\documentclass[11 pt]{article}
\usepackage[mathscr]{eucal}
\usepackage{amssymb,latexsym}
\usepackage{verbatim}
\usepackage{amsmath}
\usepackage{amsthm}
\usepackage{enumerate}
\usepackage{authblk}
\usepackage{color}
\usepackage{multicol}
\usepackage{tikz}
\usepackage{xypic}
\usepackage{caption}
\usepackage{cite}

\newtheorem{thm}{Theorem}[section]
\newtheorem{lem}[thm]{Lemma}
\newtheorem{Def}[thm]{Definition}
\newtheorem{prop}[thm]{Proposition}
\newtheorem{cor}[thm]{Corollary}

\renewcommand\l{\lambda}

\renewcommand\S{\Sigma}

\newcommand\s{\sigma}

\newcommand\e{\varepsilon}
\renewcommand\b{\beta}

\renewcommand\l{\lambda}
\newcommand\g{\gamma}

\renewcommand\a{\alpha}

\newcommand\beq{\begin{equation}}
\newcommand\eeq{\end{equation}}
\newcommand\ben{\begin{enumerate}}
\newcommand\een{\end{enumerate}}
\newcommand\bit{\begin{itemize}}
\newcommand\eit{\end{itemize}}

\newcommand{\R}{\mathbb R}

\newcommand{\ov}{\overline}

\newcommand{\pd}{\partial}

\newcommand{\mc}{\mathcal}

\def\undertilde#1{\mathord{\vtop{\ialign{##\crcr
   $\hfil\displaystyle{#1}\hfil$\crcr\noalign{\kern1.5pt\nointerlineskip}
   $\hfil\tilde{}\hfil$\crcr\noalign{\kern1.5pt}}}}}

\newcounter{mnotecount}

\title{Aspects of $C^0$ causal theory}
\author{Eric Ling\footnote{eling@math.miami.edu}}
\affil{Department of Mathematics
\\ KTH Royal Institute of Technology }

\begin{document}
\date{}
\maketitle
\vspace{.2in}

\begin{abstract}
 This paper serves as an introduction to $C^0$ causal theory. We focus on those parts of the theory which have proven useful for establishing spacetime inextendibility results in low regularity -- a question which is motivated by the strong cosmic censorship conjecture in general relativity. This paper is self-contained; prior knowledge of causal theory is not assumed.
\end{abstract}

\newpage

\tableofcontents

\newpage

\section{Introduction}

Recently, there has been an interest in low regularity aspects of Lorentzian geometry \linebreak motivated in part by the \emph{strong cosmic censorship conjecture} in general relativity. Roughly, the conjecture states that the maximal globally hyperbolic development of generic initial data for the Einstein equations is inextendible as a suitably regular Lorentzian manifold. Formulating a precise statement of the conjecture is itself a challenge since one needs to make precise the phrases `generic initial data' and `suitably regular Lorentzian manifold.' Understanding the latter is where general relativity in low regularity  and  inextendibility results become significant. 

The strongest form of the conjecture would prove inextendibility in the lowest regularity \linebreak possible -- continuity of the metric. Proving $C^0$-inextendibility results is a nontrivial \linebreak pursuit. The  classical  arguments of diverging curvature quantities only prove $C^2$-inextendibility (since the  curvature tensor requires two derivatives of the metric to be defined).  The first example of a $C^0$-inextendibility result is Sbierski's impressive proof of the $C^0$-inextendibility of Schwarzschild \cite{Sbierski, Sbierski2}.  Since then other inextendibility results have been found \cite{GallowayLing, GLS, ChrusKling, GrafLing}
and also within the context of Lorentzian length spaces \cite{GrantKunzClemens} and Lorentz-Finsler spaces \cite{MinguzziSuhr}. 

Understanding which spacetimes are $C^0$-inextendible is a highly investigated research problem in Lorentzian geometry. Therefore an understanding of causal theory for $C^0$ spacetimes is necessary for anyone who wants to break into the field. This paper serves as an introduction to $C^0$ causal theory.  We focus on those parts of the theory which have proven useful for inextendibility results. These are

\begin{itemize}
\item[-] $I^+(p)$ is open (Theorem \ref{open thm}).

\item[-] The existence of limit curves (Theorem \ref{limit curve thm}).

\item[-] The existence of causal maximizers in globally hyperbolic spacetimes (Theorem \ref{GH maximizer thm}).

\item[-] Cauchy surfaces imply global hyperbolicity (Theorem \ref{CS implies GH thm}).
\end{itemize}

The main difference between $C^0$ and smooth (at least $C^2$) causal theory is the existence of bubbling sets in $C^0$ spacetimes. This was shown in \cite{ChrusGrant}. Bubbling sets are open sets of the form $B^+(p) = \text{int}\big[J^+(p)\big] \setminus \ov{I^+(p)}$. In appendix \ref{C2 appendix} we show that $B^+(p) = \emptyset$ for all points in a $C^2$ spacetime. Hence bubbling sets are irrelevant in $C^2$ causal theory. But they play a prominent role in $C^0$ causal theory. Section \ref{bubbling set section} introduces them. In section \ref{trapped set section} we offer a notion of a trapped set for $C^0$ spacetimes and prove a $C^0$ version of Penrose's theorem: if a $C^0$ spacetime has a noncompact Cauchy surface, then there are no trapped sets.

The treatment of $C^0$ causal theory in \cite{ChrusGrant} and \cite{Clemens} uses a sequence of wider and narrower smooth metrics to approximate the $C^0$ metric. They then infer $C^0$ causal theory results from knowledge of smooth causal theory. Our approach is different. We obtain our results directly by using continuity to locally approximate the metric with wider and narrower metrics built from the Minkowski metric (Lemma \ref{coord lem}). See also \cite{MinguzziCone} which includes the results above and also generalizes causal theory even further with notable applications.

This paper is self-contained; prior knowledge of causal theory is not assumed. We only assume the Hopf-Rinow theorem  and basic integration theory.

\newpage

\section{Preliminary causal theory for $C^0$ spacetimes}

\subsection{$C^0$ spacetimes}

Let $k \geq 0$ be an integer. A $C^{k}$ \emph{metric} on a smooth manifold $M$ is a nondegenerate symmetric tensor $g\colon TM \times TM \to \R$ with constant signature whose components $g_{\mu\nu} = g(\pd_\mu, \pd_\nu)$ in any coordinate system are $C^{k}$ functions. Symmetric means $g(X,Y) = g(Y,X)$ for all $X,Y \in TM$. Nondegenerate means $g(X,Y) = 0$ for all $Y \in TM$ implies $X = 0$. With constant signature means there is an integer $r$ such that at each point $p \in M$, there is a basis $e_0, \dotsc,e_r,\dotsc, e_n \in T_pM$ such that $g(e_\mu, e_\mu) =  -1$ for $0 \leq \mu \leq r$ and $g(e_\mu, e_\mu) = 1$ for $r + 1 \leq \mu \leq n$ and $g(e_\mu, e_\nu) = 0$ for $\mu \neq \nu$. If $g(e_0, e_0) = -1 $ and $g(e_i, e_i) = 1$ for all $i = 1, \dotsc, n$, then $g$ is called a \emph{Lorentzian} metric and $(M,g)$ is called a Lorentzian manifold. If $g(e_\mu, e_\mu) = 1$ for all $\mu = 0, 1, \dotsc, n$, then $g$ is called a \emph{Riemannian} metric and $(M,g)$ is called a Riemannian manifold. Our convention will be that Greek indices  $\mu$ and $\nu$ will run through $0, 1, \dotsc, n$ and Latin indices $i$ and $j$ will run through $1, \dotsc, n$. 

If $(M,g)$ is a Lorentzian manifold, then a nonzero vector $X \in T_pM$ is \emph{timelike}, \emph{null}, or \emph{spacelike} if $g(X,X) < 0, \, =0, \,\, > 0$, respectively. A nonzero vector is \emph{causal} if it is either timelike or null. A Lorentzian manifold $(M,g)$ is \emph{time-oriented} provided there is a $C^1$ timelike vector field $X$ on $M$. A causal vector $Y \in T_pM$ is \emph{future-directed} if $g(X,Y) < 0$ and \emph{past-directed} if $g(X,Y) > 0$. Note that $-X$ defines an opposite time-orientation, and so any statement/theorem in a spacetime which is time-oriented by $X$ has a time-dual statement/theorem with respect to the time-orientation given by $-X$. 

\medskip

\begin{Def}
\emph{
Let $k \geq 0$. A $C^k$ \emph{spacetime} $(M,g)$ is a time-oriented Lorentzian manifold with a $C^k$ metric such that $M$ is connected, Hausdorff, and second-countable.
}
\end{Def}

\medskip

We now proceed to define timelike and causal curves. The class of curves that we consider should be sufficiently regular so that we can integrate along them but not too regular so that limit curves are not considered causal curves. The class of locally Lipschitz curves live in this Goldilocks zone.

Fix a $C^0$ spacetime $(M,g)$ and a smooth complete Riemannian metric $h$ on $M$.
Let $I \subset \R$ be an interval (i.e. any connected subset of $\R$ with nonempty interior). A \emph{locally Lipschitz} curve $\g \colon I \to M$ is a continuous function such that for any compact $K \subset I$, there is a constant $C$ such that for any $a,b \in K$, we have
\[
d_h\big(\g(a), \g(b)\big) \,\leq\, C|b - a|
\] 
where $d_h$ is the Riemannian distance function associated with $h$. Proposition \ref{locally lipschitz components prop} shows that we can integrate along locally Lipschitz curves.

\medskip
\medskip

\begin{prop}\label{locally lipschitz components prop}
If $\g \colon I \to M$ is locally Lipschitz, then the components $\g^\mu = x^\mu \circ \g$ in any coordinate system are differentiable almost everywhere and $(\g^\mu)' \in L^\infty_{\emph{\text{loc}}}$. Specifically, for any $t_0 \in I$, there is a coordinate system $\phi \colon U \to \R^{n+1}$ containing $\g(t_0)$  such that for any compact $K \subset I$ with $\g(K) \subset U$, there is a constant $C$ such that  $|(\g^\mu)'| \leq C$ almost everywhere in $K$ for each $\mu$.
\end{prop}

\medskip
\medskip
We include a discussion of locally Lipschitz curves in appendix \ref{locally Lipschitz appendix} where we prove Proposition \ref{locally lipschitz components prop}. But we mention here that Proposition \ref{locally lipschitz components prop} is an immediate consequence of Radaemacher's theorem which is a higher-dimensional generalization of the well-known fact that if $f\colon [a,b] \to \R$ has Lipschitz constant $C$, then $f$ is differentiable almost everywhere and $|f'| \leq C$ almost everywhere. A discussion of Rademacher's theorem in this setting along with references can be found in \cite{Chrus}.

\medskip
\medskip

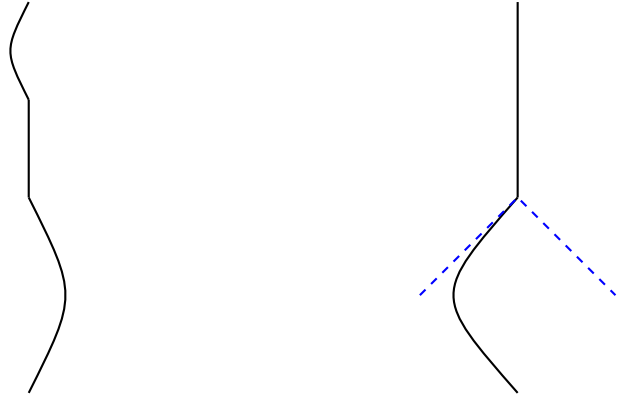
\begin{figure}[h]
\[
\begin{tikzpicture}[scale = 0.65]

\draw [thick, black] (-5,-2) .. controls (-4,0) .. (-5,2);
\draw [thick, black] (-5,2) -- (-5,4);
\draw [thick, black] (-5,4) .. controls (-5.5,5) .. (-5,6);

\draw [thick, black] (5,-2) .. controls (3.25,0) .. (5,2);
\draw[thick, black] (5,2) -- (5,6);

\draw [dashed, thick, blue](3,0) -- (5,2) -- (7,0);

\end{tikzpicture}
\]
\captionsetup{format=hang}
\caption{\small{Both of these curves are causal curves in two-dimensional Minkowski space. The curve on the left is a timelike curve. The curve on the right is a timelike almost everywhere curve. It is not a timelike curve since it approaches a null vector at its break point. }}
\label{curve fig}
\end{figure}

\begin{Def}\label{causal curve def}
\emph{
\begin{itemize}
\item[(1)] A \emph{causal curve} is a locally Lipschitz curve $\g \colon I \to M$ such that $\g'$ is future-directed causal almost everywhere.
\item[(2)] A \emph{timelike almost everywhere curve} is a causal curve $\g \colon I \to M$ such that $\g'$ is future-directed timelike almost everywhere.
\item[(3)] A \emph{timelike curve} is a causal curve $\g \colon I \to M$ such that $g(\g', \g') < -\e$ almost everywhere for some $\e > 0$. See figure \ref{curve fig}.
\end{itemize}
}
\end{Def}

\medskip
\medskip

\noindent\emph{Remarks.}
\begin{itemize}

\item[-] Our definition of a timelike curve is analogous to the locally uniform timelike curves which appear in \cite{ChrusGrant}.

 \item[-] Note that `future-directed' is implicit in our definition of causal and timelike curves. Therefore all causal and timelike curves in this paper will be future-directed.

 \item[-] A causal curve $\g$ satisfies $\g' \neq 0$ almost everywhere.

 \item[-] Given a set $S \subset M$ and a causal curve $\gamma\colon I \to M$, we will write $\gamma \subset S$ instead of $\gamma(I) \subset S$. Likewise with the intersection $\g \cap S$.
 
 \end{itemize}

\medskip
\medskip

\medskip
\medskip

Typically one constructs causal/timelike curves from smooth or piecewise smooth curves. That these are causal/timelike in the sense of Definition \ref{causal curve def} follows from the next proposition.

\medskip
\medskip

\begin{prop}\label{normal causal and timelike prop}
Let $\g \colon [a,b] \to M$ be a $C^1$ curve.  
\begin{enumerate}
\item[\emph{(1)}] If $\g'(t)$ is future-directed causal for all $t \in [a,b]$, then $\g$ is a causal curve.

\item[\emph{(2)}] If $\g'(t)$ is future-directed timelike for all $t \in [a,b]$, then $\g$ is a timelike curve.
\end{enumerate}
\end{prop}

\proof
(1) follows immediately since $C^1$ curves are locally Lipschitz (see Proposition \ref{C1 is loc Lip prop}). Now we prove (2). From (1) we know that $\g$ is a causal curve. Since $\g$ is $C^1$, we have $g(\g', \g')$ is a continuous function of $t$. Since $g(\g', \g') < 0$ and $[a,b]$ is compact, there exists an $\e > 0$ such that $g\big(\g'(t), \g'(t)\big) < -\e$ for all $t \in [a,b]$. Hence $\g$ is a timelike curve. 
\qed

\medskip
\medskip

\begin{Def}\label{causal sets definition}
\emph{
Given a set $S$ within an open neighborhood $U$, we define the \emph{causal future} and \emph{timelike future} of $S$ within $U$ as
\begin{align*}
J^+(S,U) \, &= \, \{p  \mid  \text{there is a causal curve } \g\colon [a,b] \to U\text{ with } \g(a) \in S,\, \g(b) = p \} \cup S
\\
\\
I^+(S,U) \,&=\, \{p  \mid \text{there is a timelike curve } \g\colon [a,b] \to U\text{ with } \g(a) \in S,\, \g(b) = p\}
\end{align*}
}
\end{Def}

\medskip
\medskip

\noindent \emph{Remarks.} 

\begin{itemize}

\item[-] The \emph{causal past} $J^-(S,U)$ and \emph{timelike past} $I^-(S,U)$ are defined time-dually. Any statement/theorem for $J^+$ has a corresponding time-dual statement/theorem for $J^-$. Likewise with $I^+$ and $I^-$. For example, the proof that $I^+$ is open (Theorem \ref{open thm}) has a corresponding time-dual proof that $I^-$ is open. 

\item[-] If $I^+$ is defined via timelike almost everywhere curves, then it is not necessarily open (see \cite{future_not_open}). This is the main distinction between timelike curves and timelike almost everywhere curves.

\item[-] If $U = M$, then we will write $I^+(S)$ instead. If $S = \{p\}$, then we will write $I^+(p,U)$ instead. Likewise with $J^+$. If we wish to emphasize the Lorentzian metric $g$ being used, then we will write $I^+_g$ and $J^+_g$.

\item[-] Given our convention, the constant curve $\g \colon [0,1] \to M$ given by $\g(t) = p$ for all $t$ is not a causal curve. This is why we include the union with $S$ in our definition of $J^+(S,U)$.

\end{itemize}
\medskip
\medskip

\subsection{Properties of timelike and causal curves}

\medskip
\medskip

For this section fix a $C^0$ spacetime $(M,g)$ and a complete Riemannian metric $h$ on $M$. The goal of this section is to prove the following two important properties of timelike and causal curves:

\begin{itemize}

\item[(1)] $I^+(p)$ and $I^-(p)$ are open sets. This is Theorem \ref{open thm}. 

\item[(2)] A causal curve  is inextendible if and only if it has domain $\R$ when parameterized by $h$-arclength. This is  Theorem \ref{inextend curves thm}.

\end{itemize}

\medskip

To simplify arguments, we will often parameterize causal curves by $x^0$ within a coordinate neighborhood. This is possible since $x^0$ is a \emph{time function} for a small enough neighborhood (see (3) in Lemma \ref{coord lem}). 

\medskip

\begin{Def}
\emph{ Let $U \subset M$ be open. A $C^1$ function $\tau \colon U \to \R$ is a \emph{time function} on $U$ if its gradient $\nabla \tau$ is a past-directed timelike vector field on $U$. A curve $\g \colon [a,b] \to U$ is \emph{parameterized by $\tau$} provided $\tau \circ \g(t) = t$. 
}
\end{Def}

\medskip
\medskip

\begin{prop}\label{time fun prop}
Let $\tau \colon U \to \R$ be a time function and $\g\colon [a,b] \to U$ a causal curve. Then 
\[\g \,\subset\, \big\{p \in U \mid \tau(p) \,\geq\, \tau \circ \g(a)\big\}
\]
and $\g$ has a reparameterization which is parameterized by $\tau$.
\end{prop}

\proof
Integrating gives
\[
\tau \circ \g(t) - \tau \circ \g(a) \,=\, \int_a^t (\tau \circ \g)' \,=\, \int_a^t g(\nabla \tau, \g')\, > \, 0.
\]
The last inequality holds since $\nabla \tau$ is a past-directed timelike vector field and $\g'$ is a future-directed causal almost everywhere. Moreover Proposition \ref{locally lipschitz components prop} implies that the above integral is finite. Thus $\tau \circ \g$ is a strictly increasing continuous function with a positive derivative almost everywhere; hence it is invertible with continuous inverse that is differentiable almost everywhere.
The reparameterization we seek is $\tilde{\g} = \g \circ (\tau \circ \g)^{-1}$. 
\qed

\medskip
\medskip

\begin{Def}
\emph{
The \emph{Minkowski} metric on $\R^{n+1}$ is $\eta = \eta_{\mu\nu}dx^\mu dx^\nu = -(dx^0)^2 + \delta_{ij}dx^idx^j$. For $0 < \e < 1$, we define the \emph{narrow} and \emph{wide} Minkowski metrics 
\[
\eta^\e \,=\, -\frac{1 - \e}{1+\e}(dx^0)^2 \,+\, \delta_{ij}dx^idx^j \,\,=\,\, \eta + \frac{2\e}{1+\e}(dx^0)^2
\]
\[
\eta^{-\e}  \,=\, -\frac{1 + \e}{1-\e}(dx^0)^2 \,+\, \delta_{ij}dx^idx^j \,\,=\,\,  \eta - \frac{2\e}{1-\e}(dx^0)^2
\]
}
\end{Def}

\medskip
\medskip

\noindent\emph{Remark}. For example $\eta^{3/5}$ and $\eta^{-3/5}$ have lightcones with `slopes' 2 and 1/2, respectively. Note that as $\e$ approaches $0$, we have $\eta^\e$ and $\eta^{-\e}$ approach $\eta$.

\medskip
\medskip

\begin{lem}\label{coord lem}
Fix $p \in M$. For any $0< \e < 1$, there is a coordinate system $\phi \colon U_\e \to \R^{n+1}$  with the following properties 
\begin{itemize}
\item[\emph{(1)}] $\phi(p) \, = \, 0$

\item[\emph{(2)}] $g_{\mu\nu}(p) \, =\, \eta_{\mu\nu}$ and $|g_{\mu\nu}(x) - \eta_{\mu\nu}| \,<\, \e$ for all $x \in U_\e$

\item[\emph{(3)}] $x^0$ is a time function on $U_\e$.

\end{itemize}

\noindent For all $q \in U_\e$ and nonzero $X \in T_qM$, we have

\begin{itemize}

\item[\emph{(4)}] $\eta^\e(X,X) <0 \:\, \Longrightarrow\:\, g(X,X) <0 \:\,\Longrightarrow\:\, g(X,X) \leq 0 \:\,\Longrightarrow \:\, \eta^{-\e}(X,X) < 0$

\item[\emph{(5)}] $I^+_{\eta^{\e}} (q, U_\e) \,\subset\,I^+(q, U_\e) \,\subset\, J^+ (q, U_\e) \,\subset\, I^+_{\eta^{-\e}} (q, U_\e)$.

\end{itemize}

\end{lem}

\medskip

\noindent\emph{Remark.} Technically, in (4) and (5) we should write $\phi^* \eta^\e$ instead of $\eta^{\e}$ where $\phi^*\eta^\e$ is the pullback metric, likewise with $\eta^{-\e}$. We will continue to use this abuse of notation whenever we apply Lemma \ref{coord lem}.

\medskip

\proof

Pick a coordinate system $\phi \colon U \to \R^{n+1}$ with $\phi(p) = 0$ to obtain (1) and apply Gram-Schmidt to obtain the first part of (2) with $\pd_0$ future-directed timelike at $p$. By continuity of the metric, given any $\e_0 > 0$, we can shrink our neighborhood so that $|g_{\mu\nu}(x) - \eta_{\mu\nu}| < \e_0$ for all $x \in U$. Choose $\e_0 < \e$ to obtain the second part of (2). Since $\e_0 < \e < 1$,  we have $g_{00}(x) < 0$. Therefore $\pd_0$ is future-directed timelike in $U$. Since $1 = dx^0(\pd_0) =  g(\nabla x^0, \pd_0)$, we have $\nabla x^0$ is past-directed timelike. This shows (3). 

 To show (4), let $X = X^\mu \pd_{\mu}$ be any tangent vector in $T_qU$. Then we have
\begin{align*}
g_{\mu\nu}X^\mu X^\nu \,&=\, g_{00}|X^0|^2 + \sum_{i=1}^n g_{ii}|X^i|^2 + 2\sum_{i=1}^n g_{0i}X^0X^i + \sum_{i \neq j}g_{ij}X^i X^j 
\\
&<\, (\eta_{00} + \e_0)|X^0|^2 + \sum_i (\eta_{ii} + \e_0)|X^i|^2 + 2\e_0 \sum_{i}|X^0X^i| + \e_0\sum_{i \neq j}|X^iX^j|
\\
&=\, \eta_{\mu\nu}X^\mu X^\nu + \e_0 \sum_{\mu,\nu}|X^\mu X^\nu|.
\end{align*}
Using the fact that $\eta = \eta^\e - \frac{2\e}{1+\e}(dx^0)^2$, we have
\[
g(X,X) \,<\, \eta^\e(X,X) - \frac{2\e}{1+\e}|X^0|^2 + \e_0\left[|X^0|^2 + 2\sum_i |X^0X^i| + \sum_{i,j}|X^i X^j| \right].
\]
Now suppose $X$ is $\eta^{\e}$-timelike, then $|X^i|^2/|X^0|^2 < (1-\e)/(1+\e)$ for each $i$. Therefore 
\[
g(X,X) \,<\, \eta^\e(X,X) - \frac{2\e}{1+\e}|X^0|^2 + \e_0 \left[1 + 2n \sqrt{\frac{1-\e}{1+\e}} + n^2\frac{1-\e}{1+\e} \right]|X^0|^2.
\]
By taking $\e_0 > 0$ small enough, we can ensure $2\e/(1+\e)$ is strictly greater than the bracket term. Then for this choice of $\e_0$, we have
\[
g(X,X) \,<\, \eta^{\e}(X,X) .
\]
This proves the first implication in (4). The second implication is obvious. Now we prove the third implication. Note it suffices to prove it for $X$ satisfying $h(X,X) = 1$. A similar computation as above shows that 
$\eta^{-\e}(X,X) < g(X,X) - \frac{2\e}{1-\e}|X^0|^2 + \e_0 \sum_{\mu,\nu}|X^\mu X^\nu|$. So in the same way we proved the first implication, it suffices to show that there is a neighborhood of $p$ such that $|X^i|/|X^0|$ is bounded for all $i$. Suppose this were not true for $i = 1$. Then there is a sequence of points $p_k \in U$ witih $p_k \to p$ and vectors $X_k \in T_{p_k}M$ such that $g(X_k, X_k) \leq 0$, $h(X_k, X_k) = 1$, and $|X^1_k|/|X^0_k| \to \infty$. Choosing $U$ to have compact closure, the $h$-unit bundle of $U$ within $TM$ has compact closure.
 Therefore there is a subsequence (still denoted by $X_k$) such that $X_k \to X_* \in T_pM$ and continuity implies $h(X_*,X_*) = 1$ and $g(X_*,X_*) \leq 0$. The former implies $X_* \neq 0$ and so the latter implies $X_*$ is a causal vector.  But the limit $|X^1_k|/|X^0_k| \to \infty$ implies $X^0_* = 0$ which implies $X_*$ is spacelike -- a contradiction. Therefore there is a neighborhood of $p$ such that $|X^1|/|X^0|$ is bounded. Similarly this holds for all $i = 2, \dotsc, n$. Hence there is a neighborhood of $p$ such that $|X^i|/|X^0|$ is bounded for all $i$. This proves the third implication in (4). 

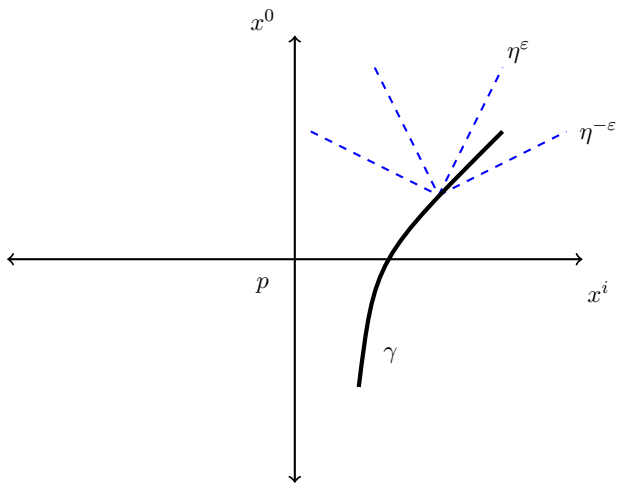
\begin{figure}[h]
\[
\begin{tikzpicture}[scale = 0.85]

\draw [<->,thick] (0,-3.5) -- (0,3.5);
\draw [<->,thick] (-4.5,0) -- (4.5,0);

\draw (-0.5,3.75) node [scale = .85] {$x^0$};
\draw (4.75,-0.5) node [scale = .85] {$x^i$};

\draw (-0.5,-0.4) node [scale = .85] {$p$};


\draw [dashed, thick, blue](0.25,2) -- (2.25,1) -- (4.25,2);
\draw [dashed, thick, blue](1.25,3) -- (2.25,1) -- (3.25,3);

\draw (3.5,3.25) node [scale = .85] {$\eta^\e$};
\draw (4.75,2) node [scale = .85] {$\eta^{-\e}$};


\draw [ultra thick, black] (1,-2) .. controls (1.25,0) .. (3.25,2);

\draw (1.5,-1.5) node [scale = .85] {$\gamma$};


\end{tikzpicture}
\]
\captionsetup{format=hang}
\caption{\small{The coordinate system $\phi \colon U_\e \to \R^{n+1}$ appearing in Lemma \ref{coord lem}. The point $p$ is located at the origin where the metric is exactly Minkowski: $g_{\mu\nu}(p) = \eta_{\mu\nu}$. Any causal curve $\g \subset U_\e$ will always be $\eta^{-\e}$-timelike but it may be $\eta^\e$-spacelike.}}
\label{coord fig}
\end{figure}

Now we prove (5). Let $\g\colon [a,b] \to U$ be an $\eta^\e$-timelike curve. Then there is a $\delta > 0$ such that $\eta^\e(\g',\g') < -\delta$ almost everywhere. Then by the first implication in (4) we have $g(\g', \g') < \eta^\e(\g', \g') < -\delta$ almost everywhere. Hence $\g$ is a $g$-timelike curve. This shows the first inclusion in (5). The second inclusion is obvious. To prove the third inclusion, recall from the proof of the third implication in (4), we have for any tangent vector \linebreak $\eta^{-\e}(X,X) < g(X,X) - \frac{2\e}{1-\e}|X^0|^2 + \e_0 \sum_{\mu,\nu}|X^\mu X^\nu|$. Recall we chose $\e_0$ sufficiently small so that $\frac{2\e}{1-\e}|X^0|^2$ was strictly greater than $\sum_{\mu,\nu}|X^\mu X^\nu|$. Hence there is a $\delta > 0$ such that $\eta^{-\e}(X,X)< g(X,X) - \delta$. Thus, if $\g$ is $g$-causal, then $\g$ is $\eta^{-\e}$-timelike. See figure \ref{coord fig}.
\qed
%

\medskip
\medskip

%
%
%
%
%
%

%
%

\medskip
\medskip

\begin{Def}\label{cone def}
\emph{
Let $\langle \cdot, \cdot \rangle$ and $|\cdot|$ denote the standard inner product and norm on $\R^{n+1}$ with its standard global orthonormal basis $\{e_0, e_1, \dotsc, e_n\}$. Given any open set $U \subset \R^{n+1}$ and any point $p \in U$, we define for $-1 < \e < 1$
\[
C^+_\e (p,U) \,=\, \left\{ q \in U \: \bigg| \: \frac{\langle q - p,\, e_0\rangle}{|q - p|} \,>\, \sqrt{\frac{1+\e}{2}}\right\}.
\]
}
\end{Def}

\medskip
\medskip

\noindent \emph{Remark.} $C^+_\e$ is the usual interior of a cone in $\R^{n+1}$ which makes an angle $\theta$ with respect to the $x^0$-axis where $\theta$ is given by $\cos\theta = \sqrt{(1+\e)/2}$. Note that $C^+_0$ coincides with the future lightcone in Minkowski space. 

\medskip
\medskip

\begin{lem}\label{euclidean cone lem}
Let $B$ be an open ball in $\R^{n+1}$ and let $p \in B$. Then 
\begin{itemize}

\item[\emph{(1)}] $C^+_\e(p, B) \, \subset\, I^+_{\eta^\e}(p, B)\,\subset\, C^+_{\e'}(p, B)$ \:\:\:\:\:\:\:\:\:\:\:\:\:\:\:\:  $0< \e' < \e < 1$

\item[\emph{(2)}] $C^+_{-\e}(p, B) \, \subset\, I^+_{\eta^{-\e}}(p, B)\,\subset\, C^+_{-\e'}(p, B)$ \:\:\:\:\:\:\:\:\:\:  $0< \e < \e' < 1$

\end{itemize}

%

\end{lem}

\proof
We only prove (1) as the proof of (2) is analogous.

 We first prove the left inclusion of (1). Let $q \in C^+_\e(p, B)$. Let $\g\colon [0,1] \to B$ be the straight line $\g(t) = qt + (1-t)p$. Then $\g'(t) = q-p$. Put $q - p = X = X^\mu e_\mu$. By definition we have $X^0/|X| > \sqrt{(1+\e)/2}$. Notice that $|X|^2 = \langle X, \, X \rangle = |X^0|^2 + \delta_{ij}X^iX^j$. Hence $|X^0|^2 > \frac{1}{2}(1+\e)\big(|X^0|^2 + \delta_{ij}X^iX^j\big)$. Rearranging gives $|X^0|^2 > \frac{1+\e}{1-\e}\delta_{ij}X^iX^j$. Therefore
 \[
 \eta^\e(\g', \g') \,=\, \eta^\e(X,X) \,=\, -\frac{1 - \e}{1+\e}|X^0|^2 \,+\, \delta_{ij}X^iX^j \,<\, 0.
 \]
Therefore $\g$ is an $\eta^\e$-timelike curve by  Proposition \ref{normal causal and timelike prop}. Hence $q \in I^+_{\eta^\e}(p,B)$. 
 
 Now we prove the right inclusion of (1). Suppose $q \in I^+_{\eta^\e}(p,B)$. Let $\gamma\subset B$ be an $\eta^\e$-timelike curve from $p$ to $q$. To help visualize the proof consider $\e = 15/17$ and $\e' = 3/5$ which correspond to lightcones with `slopes' 4 and 2, respectively. Consider the hyperplanes given by $x^0 - 2x^1 = \text{constant}$. Note that these hyperplanes are $\eta^{\e'}$-null but $\eta^\e$-spacelike. Let $\tau$ be the $\eta^{\e}$-time function such that $\nabla \tau$ is orthogonal to these hyperplanes. Apply Proposition \ref{time fun prop} with $g = \eta^\e$ to conclude that $\g$ lies above the particular hyperplane which intersects $p$. Now replace $x^1$ with any arbitrary direction orthogonal to $\pd/ \pd x^0$, and apply Proposition \ref{time fun prop} again to conclude that  $\g \subset  C^+_{\e'}(p,B).$ Clearly this proof does not depend on the specific choices of $\e$ and $\e'$.
\qed

\medskip
\medskip

\begin{thm}\label{open thm}
$I^+(p,U)$ is open.
\end{thm}

\proof
Fix $q \in I^+(p, U)$. Let $\g \colon [a,b] \to U$ be a timelike curve from $p$ to $q$. Choose a coordinate system $\psi \colon V \to \R^{n+1}$  around $\g(b)$ from  Proposition \ref{locally lipschitz components prop} such that $V \subset U$ and $V$ has compact closure. By continuity, there is an $a_0 \in (a,b)$ such that $\g|_{[a_0,b]} \subset V$. Since $[a_0,b]$ is compact, there is a constant $C$ such that $|(y^\mu \circ \g)'| \leq C$ almost everywhere in $[a_0, b]$ where $(y^0, y^1, \dotsc, y^n)$ are the coordinates on $V$. 

 Let $\phi \colon U_\e \to \R^{n+1}$ be a coordinate system about $q$ from Lemma \ref{coord lem} and choose $U_\e$ small enough so that $U_\e \subset V$. Let $(x^0, x^1, \dotsc, x^n)$ denote the coordinates on $U_\e$. Let $X = \g'$. By definition of a timelike curve, there is a $\delta > 0$ such that $g(X, X) < -\delta$  almost everywhere. For the portion of $\g$ within $U_\e$, write $X = X^\mu \pd/\pd x^\mu$. Using $\eta_{\mu\nu} < g_{\mu\nu}(x) + \e$ and a similar calculation as in the proof of Lemma \ref{coord lem}, we have
\[
\eta(X,X) -  \e \sum_{\mu, \, \nu}|X^\mu X^\nu| \, < \, g(X,X) \, < \, - \delta.
\]
Since $\eta = \eta^\e - \frac{2\e}{1 + \e}(dx^0)^2$, we have
\[
\eta^\e(X,X) - \frac{2\e}{1+\e}|X^0|^2 - \e\sum_{\mu, \, \nu}|X^\mu X^\nu| \, < \, -\delta.
\]
Rearranging gives
\[
\eta^\e(X,X) \,<\, -\delta + \frac{2\e}{1+\e}|X^0|^2 + \e \left[1 + 2\sum^n_{i=1} \frac{|X^i|}{|X^0|} + \sum_{1 \leq i, \, j \leq n} \frac{|X^i X^j|}{|X^0|^2} \right]|X^0|^2.
\]
Choose $\e < 3/5$. Then by the third implication in (4) of Lemma \ref{coord lem}, we have $|X^i|/|X^0| < 2$. Therefore the term in the bracket is bounded by $1 + 4n + 4n^2$, and so we have
\[
\eta^\e(X,X) \,<\, -\delta + \frac{2\e}{1+\e}|X^0|^2 + \e\big(1 + 4n + 4n^2\big) |X^0|^2.
\] 
Now we put a bound on $|X^0|$. We have $|X^0| = |(y^\mu \circ \g)' \pd x^0/ \pd y^\mu| \leq C \sum_\mu| \pd x^0/ \pd y^\mu|$. The inequality follows from the first paragraph of this proof. Since $V$ has compact closure, there exists a $c$ such that $|\pd x^0/\pd y^\mu| \leq c$ for all points in $U_\e$. Thus $|X^0| \leq Cc(n+1)$.  

Therefore we can choose $\e$ sufficiently small such that
\[
\eta^\e(X,X) \,<\, -\frac{1}{2}\delta.
\]
Hence, by shrinking $U_\e$ small enough, the portion of $\g$ within $U_\e$ is $\eta^\e$-timelike. Let $B \subset U_\e$ be a neighborhood around $q$ such that $\phi(B)$ is an open Euclidean ball centered around $\phi(q)$. By continuity, there exists an $a_1 < b$ such that $ \g|_{[a_1, b]} \subset B$. Let $p_1 = \g(a_1)$ and recall $q = \g(b)$. Then we just showed $q \in I^+_{\eta^\e}\big(p_1,\, B\big)$. Choose $\e' \in (0,\e)$. Then (1) from Lemma \ref{euclidean cone lem} implies
\[
q \,\in\, \phi^{-1} \circ C^+_{\e'}\big(\phi(p_1),\, \phi(B)\big) \,\subset\, I^+_{\eta^{\e'}}(p_1,\, B) \,\subset\, I^+(p_1,\, B).
\]
The last inclusion follows from (5) in Lemma \ref{coord lem}. Since $C^+_{\e'}\big(\phi(p_1), \phi(B)\big)$ is an open set and its preimage is contained in $I^+(p_1, \, B) \subset I^+(p,U)$, we have $I^+(p,U)$ is open.
\qed


\medskip
\medskip

\begin{cor}
$I^+(S,U)$ is open.
\end{cor}

\proof
$I^+(S,U) = \bigcup_{p \in S}I^+(p,U)$.
\qed

\medskip
\medskip

Now we set out to prove the second claim of this section: causal curves are inextendible when they have domain $\R$ when parameterized by $h$-arclength. A locally Lipschitz curve $\g \colon I \to M$ is parameterized by \emph{$h$-arclength} if $h(\g', \g') = 1$ almost everywhere. The next proposition will allow us to parameterize causal curves with respect to $h$-arclength.

\medskip
\medskip

\begin{prop}\label{arclength prop}
Let $\g \colon I \to M$ be a causal curve. Then $\g$ admits a reparameterization $\tilde{\g}$ such that $\tilde{\g}$ is parameterized by $h$-arclength and for all $a, b$ in the domain of $\tilde{\g}$, we have
\[
d_h\big(\tilde{\g}(a), \tilde{\g}(b)\big) \,\leq\, |a - b|.
\]
\end{prop}

\proof
By Proposition \ref{locally lipschitz components prop}, the components of $\g$ are differentiable almost everywhere and these derivatives are in $L^\infty_{\text{loc}}$. Therefore the integral 
\[
s(t) \,=\, \int_{t_0}^t\sqrt{h(\g', \g')}
\]
is well-defined and finite where $t_0, t \in I$. Since $\g$ is a causal curve, we have $\g' \neq 0$ almost everywhere. Therefore $s(t)$ is a strictly increasing continuous function; hence it is invertible. Moreover $s$ is differentiable almost  everywhere with $s' > 0$ wherever differentiable; hence $s^{-1}$ is differentiable almost everywhere.  The reparameterization we seek is $\tilde{\g} = \g \circ s^{-1}$. Then in any coordinate system, we have $h_{\mu\nu}(\tilde{\g}^\mu)'(\tilde{\g}^\nu) ' = 1$.  Hence $h(\tilde{\g}', \tilde{\g}') = 1$ almost everywhere. Thus
\[
\pushQED{\qed}
b - a \,=\, \int_a^b dt \,=\, \int_a^b \sqrt{h(\tilde{\g}', \tilde{\g}')} \, \geq \, d_h\big(\tilde{\g}(a), \tilde{\g}(b)\big). \qedhere
\popQED
\]

\medskip
\medskip

\begin{Def}
\emph{
Let $\g \colon [a,b) \to M$ be a causal curve. Suppose there exists a $p \in M$ such that $\g(t_n) \to p$ for every sequence $t_n \nearrow b$. Then $p$ is called the \emph{future endpoint} of $\g$. \emph{Past endpoints} are defined time-dually.
}
\end{Def}

\medskip
\medskip

\noindent\emph{Remark.} Future and past endpoints are unique since $M$ is Hausdorff.

\medskip
\medskip

  When $\g \colon [a,b) \to M$ has a future endpoint $p$, one is tempted to define a new curve $\tilde{\g} \colon [a,b] \to M$ such that $\tilde{\g}(t) = \g(t)$ for $t < b$ and $\tilde{\g}(b) = p$. However a problem arises: the extended curve $\tilde{\g}$ may not be locally Lipschitz. For example if one extends the curve $t \mapsto \big(\sqrt{t+1}, 0\big)$ in two-dimensional Minkowski space from $(-1,0]$ to $[-1,0]$ so that it includes the past endpoint $(0,0)$, then the new curve defined on $[-1,0]$ will not be locally Lipschitz since $\big(\sqrt{t + 1} - \sqrt{t' + 1}\,\big)/(t - t')$ diverges as $t$ and $t'$ approach $0$. However if we reparameterize causal curves with respect to $h$-arclength, then this problem goes away.

\medskip
\medskip

\begin{prop}\label{endpoint prop}
Let $\g\colon [0,b) \to M$ be a causal curve parameterized by $h$-arclength. If $b < \infty$, then there is a future endpoint $p$ of $\g$ and the curve $\tilde{\g} \colon [0,b] \to M$ defined by $\tilde{\g}(t) = \g(t)$ for $0 \leq t <b$ and $\tilde{\g}(b) = p$ is a causal curve. 
\end{prop}

\proof

Since $\g$ is parameterized with respect to $h$-arclength, we have $b = L_h(\g)$. Therefore 
\[
\g \, \subset  \,B_h\big(\g(a),\, b+1\big)
\]
where $B_h$ denotes the $h$-geodesic ball. Write $B = B_h\big(\g(a),\, b+1\big)$. Then $\ov{B}$ is compact by the Hopf-Rinow theorem. Let $t_n \nearrow b$ be any sequence. Then $\g(t_n)$ has an accumulation point $p \in \ov{B}$. Therefore there is a subsequence (still denoted by $t_n$) such that $t_n \nearrow b$ and $\g(t_n) \to p$. We show that $p$ is the future endpoint of $\g$. Let $\g(s_n)$ be any other sequence with $s_n \nearrow b$. Then $d_h\big(\g(s_n), \g(s_m)\big) \leq |s_n - s_m|$ implies that $\g(s_n)$ is a Cauchy sequence. Since $(M,h)$ is complete, the Hopf-Rinow theorem implies $\g(s_n)$ converges to some point $q$. Put $p_n = \g(t_n)$ and $q_n = \g(s_n)$. Then the triangle inequality gives
\[
d_h(p,q) \,\leq\, d_h(p,p_n) + d_h(p_n, q_n) + d_h(q_n, q).
\]
Each of the three terms on the right hand side can be made arbitrarily small. Therefore $d_h(p,q) = 0$ and so $p = q$. Hence $p$ is the future endpoint of $\g$. 

Define the continuous function $\tilde{\g} \colon [0,b] \to M$ by $\tilde{\g}(t) = \g(t)$ for $0 \leq t <b$ and $\tilde{\g}(b) = p$. Since $\g$ is parameterized by $h$-arclength, Proposition \ref{arclength prop} implies
\[
d_h\big(\g(t), \g(t')\big) \,\leq \, |t - t'|.
\]
Passing to the limit $t' \nearrow b$, continuity of $d_h$ implies
\[
d_h\big(\g(t), p\big) \, \leq \, |t - b|.
\]
Hence $\tilde{\g}$ is Lipschitz with Lipschitz constant 1. Therefore $\tilde{\g}$ is a causal curve.
\qed

\medskip
\medskip

The following technical proposition is needed for Theorem \ref{inextend curves thm}.

\medskip

\begin{prop}\label{causal bounded by h prop}
Given any $p \in M$ and $\e > 0$, there is a neighborhood $U$ such that $L_h(\g) < \e$ for all causal curves $\g \subset U$. 
\end{prop}

\proof  Fix $\e > 0$. Choose a neighborhood $\phi \colon U_{3/5} \to \R^{n+1}$ as in Lemma \ref{coord lem}. Then the lightcones of $\eta^{-3/5}$ have `slope' 1/2. Shrink $U_{3/5}$ so that it has compact closure and $-\e < x^0 < \e$. Let $\g \subset U_{3/5}$ be any causal curve. By (3) from Lemma \ref{coord lem} we can assume $\g$ is parameterized by the time function $x^0$. Put $X = \g'$ so that $X^0 = 1$. Therefore
\[
h(\g',\g') \,=\, h_{\mu\nu}X^\mu X^\nu \,=\,   h_{00} + 2 h_{0i}X^i + h_{ij}X^i X^j.
\]
Set $H = \sup \big\{|h_{\mu\nu}(q)|  \mid  q \in U_{3/5} \text{ and } 0 \leq \mu, \nu \leq n\big\}$. Then $H < \infty$ since $U_{3/5}$ has compact closure. By the third implication in (4) of Lemma \ref{coord lem}, we have $|X^i| < 2$ for each $i$. Therefore $h(\g', \g') < HC$ where $C = 1 + 4n + 4n^2$.  Since  $\g$ is parameterized by $x^0$, we have $L_h(\g) = \int  \sqrt{h(\g', \g')}dx^0\leq 2\e\sqrt{HC}$. Since $\e> 0$ was arbitrary, the result follows. 
\qed

\medskip
\medskip

\begin{Def}
\emph{
Let $\g \colon [a,b) \to M$ be a causal curve. We say $\g$ is \emph{future extendible} if there is  a causal curve  $\tilde{\g} \colon [a,b] \to M$ with $\tilde{\g} (t) = \g(t)$ for all $t \in [a,b)$. Otherwise $\g$ is \emph{future inextendible}. Time-dualizing gives  \emph{past inextendible} causal curves. $\g$ is \emph{inextendible} if it is both future and past inextendible. 
}
\end{Def}

\medskip
\medskip

\noindent\emph{Remark.} Proposition \ref{endpoint prop} shows that if $p$ is a future endpoint for a causal curve $\g$ parameterized by $h$-arclength, then $\g$ is future extendible.

\medskip
\medskip

\newpage

\begin{thm}\label{inextend curves thm}
Let $\g \colon [0, b) \to M$ a causal curve parameterized by $h$-arclength. 
\begin{itemize}
\item[\emph{(1)}] If $b = \infty$, then $\g$ is future inextendible.
\item[\emph{(2)}] If $b < \infty$, then $\g$ can be extended to a future inextendible causal curve.
\end{itemize}
\end{thm}

\proof We first prove (1). Seeking a contradiction, suppose $\g$ is future extendible. Then there is a causal curve $\tilde{\g} \colon [0, \infty] \to M$ which extends $\g$. Let $p = \tilde{\g}(\infty)$. Let $U$ be any neighborhood of $p$. By continuity, there exists a $c > 0$ such that $\tilde{\g}|_{[c,\infty]} \subset U$. But $L_h\big(\tilde{\g}|_{[c,\infty]}\big) = \infty$ which contradicts Proposition \ref{causal bounded by h prop}. This proves (1).

Now we prove (2). Suppose $b < \infty$ and set 
\[
c = \sup \big\{c' > b  \mid \g \text{ extends to a $h$-arclength parameterized causal curve on } [a,c') \big\}.
\]
By Proposition \ref{endpoint prop}, we know that $\g$ extends to a causal curve $\tilde{\g}$ on $[0,b]$. Using a coordinate system from Lemma \ref{coord lem} centered around $\tilde{\g}(b)$, we can extend $\tilde{\g}$ even further (e.g. by concatenating $\tilde{\g}$ with the positive $x^0$-axis). Therefore the set appearing in the above supremum is nonempty. Suppose $c < \infty$. Then  there is a causal curve $\lambda \colon [0,c) \to M$ which extends $\g$. Proposition \ref{endpoint prop} implies $\l$ extends to a causal curve $\tilde{\l}\colon [0,c] \to M$. We can extend $\tilde{\l}$ using a coordinate neighborhood via the same argument above. This contradicts the definition of $c$. Therefore $c = \infty$. Thus $\g$ extends to a $h$-arclength parameterized causal curve on $[0, \infty)$ which is future inextendible by (1). This proves (2). 
\qed

\medskip

\subsection{Limit curves}

Fix a $C^0$ spacetime $(M,g)$ with a complete Riemannian metric $h$ on $M$. The purpose of this section is to prove the limit curve theorem in the $C^0$ setting.

\medskip

\begin{Def}
\emph{
Let $\g_n\colon I \to M$ be a sequence of causal curves. A causal curve $\g \colon I \to M$ is a \emph{limit curve} of $\g_n$ if there is a subsequence of $\g_n$ which converges to $\g$ uniformly on compact subsets of $I$.
}
\end{Def}

\medskip
\medskip

\noindent\emph{Remark.} Limit curves are not necessarily unique. For this reason limit curves are called `accumulation curves' in \cite{Chrus, ChrusGrant}.

\medskip
\medskip

Let $\g_n \colon I \to M$ be a sequence of causal curves. We say $\g_n$ \emph{accumulates} to $p$ at $t_0$ if there is a subsequence $\g_{n_k}$ such that $\g_{n_k}(t_0) \to p$.

\medskip
\medskip

\begin{thm}[Limit Curve Theorem]\label{limit curve thm}
Let $\g_n \colon \R \to M$ be a sequence of causal curves parameterized by $h$-arclength. If $\g_n$ accumulates to $p$ at $t_0$, then there is an inextendible limit curve $\g\colon \R \to M$ of $\g_n$ such that $\g(t_0) = p$.
\end{thm}

\medskip
\medskip

The proof of of the limit curve theorem requires three arguments: (1) An argument proving the existence of the limit curve $\g$ which will follow from the Arzel{\'a}-Ascoli theorem. (2) An argument to prove that $\g$ is in fact a causal curve. (3) An argument showing that $\g$ is indeed inextendible. This last part is necessary, since, although each $\g_n$ is parameterized by $h$-arclength, there is no guarantee that the limit curve $\g$ will be.

\medskip

\begin{thm}[Arzel{\'a}-Ascoli]
Let $(M,d)$ be a metric space. If the sequence $\g_n \colon \R \to M$ is equicontinuous and for each $t \in \R$ the set $\bigcup_n \{\g_n(t)\}$ is bounded, then there exists a continuous $\g \colon \R \to M$ and a subsequence of $\g_n$ which converges to $\g$ uniformly on compact subsets of $\R$. 
\end{thm}

\medskip

\begin{prop}\label{existence limit curve prop}
Let $\g_n \colon \R \to M$ be a sequence of causal curves parameterized by $h$-arclength. If  $\g_n$ accumulates to $p$ at $t_0$, then there is a locally Lipschitz curve $\g \colon \R \to M$ with $\g(t_0) = p$ and a subsequence of $\g_n$ which converges to $\g$ uniformly on compact subsets of $\R$. 
\end{prop}

\proof
We apply the Arzel{\'a}-Ascoli theorem to the metric space $(M, d_h)$. Equicontinuity follows from the $h$-arclength parameterization: For $\e > 0$ choose $\delta = \e$. Then for all $n$ and all $|a - b| < \delta$, we have
\[
d_h\big(\g_n(a), \g_n(b)\big) \,\leq\, |a-b|  \,<\, \e.
\]
By assumption there is a subsequence (still denoted by $\g_n$) such that $\g_n(t_0) \to p$. By restricting to a further subsequence (still denoted by $\g_n$), we can assume $d_h\big(\g_n(t_0), p\big) < 1$ for all $n$. 
Then for each $n$ we have $\g_n(t) \in \{q \mid d_h(p,q) \leq |t - t_0| + 1\}$. This set is bounded by $2|t-t_0| + 2$, and so $\bigcup_n\{\g_n(t)\}$ is bounded. Thus the Arzel{\'a}-Ascoli theorem applies to this subsequence.
Therefore there is a continuous curve $\g \colon \R \to M$ and a subsequence (still denoted by $\g_n)$ which converges to $\g$ uniformly on compact subsets of $\R$. Hence $\g(t_0) = p$. That $\g$ is in fact Lipschitz (and hence locally Lipschitz) follows since $d_h$ is continuous:
\[
\pushQED{\qed}
d_h\big(\g_n(a), \g_n(b)\big) \,\leq\, |a-b| \:\:\:\:\:\: \longrightarrow \:\:\:\:\:\:
d_h\big(\g(a), \g(b)\big) \,\leq\, |a - b|. \qedhere
\popQED
\]

\medskip
\medskip

\begin{lem}\label{causal limit curve diff point}
Suppose $\g_n \colon I \to M$ is a sequence of causal curves which converges uniformly to a locally Lipschitz curve $\g \colon I \to M$ on compact subsets of $I$. If $t_0$ is a differentiable point of $\g$ and $\g'(t_0) \neq 0$, then $\g'(t_0)$ is future-directed causal.
\end{lem}
\proof
Seeking a contradiction, suppose $\g'(t_0)$ is not future-directed causal. Since $\g'(t_0) \neq 0$, it is either spacelike or past-directed causal. First suppose $\g'(t_0)$ is spacelike.  Without loss of generality, assume $t_0 = 0$. Set $p = \g(0)$ and $X = \g'(0)$.   Let $\phi \colon U_{3/5} \to \R^{n+1}$ be a coordinate system as in Lemma \ref{coord lem} centered around $p$ and apply the Gram-Schmdit process so that $X = X^1\pd/\pd x^1 |_p$ with $X^1 > 0$. Then the definition of the derivative gives $\gamma^1(t)/t \to X^1$ and $\gamma^\mu(t)/t \to 0$ for $\mu \neq 1$ as $t \to 0$. 

Given any $\e > 0$, there exists a $\delta > 0$ such that $|\g^0(t)/t| < \e$ and $|\g^1(t)/t -X^1| < \e$ for all $0 < t < \delta$. Hence $\g^0(t) / \g^1(t) < \e/(X^1 - \e)$  for  these $t$. Fix $\e$ sufficiently small so that $\g^0(t)/\g^1(t) < 1/4$ for all $0 < t < \delta$. Fix $t_1 \in (0,\delta$). Let $q \in U_{3/5}$ be a point on the negative $x^0$-axis. Write $V_q = I^+_{\eta^{-3/5}}(q,U_{3/5})$. Then $V_q$ is a neighborhood of $p$. Therefore for sufficiently large $n$, we have $\g_n(0) \in  V_q$. Since $\g_n$ converges uniformly to $\g$ on $[0,t_1]$, we have $\g_n(0) \to \g(0)$ and $\g_n(t_1) \to \g(t_1)$.  Since $\eta^{-3/5}$ has lightcones with `slope' 1/2 and $\g^0(t_1)/\g^1(t_1) < 1/4$, by choosing $q$ sufficiently close to $p$, the convergence $\g_n(t_1) \to \g(t_1)$ contradicts the second inclusion of (2) from Lemma \ref{euclidean cone lem}.  

Similarly, assuming $\g'(t_0)$ is past-directed causal leads to a contradiction.
\qed

%
%
%

\medskip
\medskip

\begin{prop}\label{causal limit curve prop}
Suppose $\g_n \colon I \to M$ is a sequence of causal curves parameterized by $h$-arclength which converges uniformly to a locally Lipschitz curve $\g \colon I \to M$ on compact subsets of $I$. Then $\g$ is a causal curve.
\end{prop}

\proof
By Lemma \ref{causal limit curve diff point} it suffices to show that $\g' \neq 0$ almost everywhere. Let $t_0 \in I$ and $p = \g(t_0)$. Consider a neighborhood $U$ from Lemma \ref{coord lem} centered around $p$. Assume $U$ has compact closure. Define $C$ by
\[
C \,=\, \inf \,\{g(\nabla x^0, X) \mid q \in U, \:X \in T_qM, \: g(\nabla x^0, X) > 0,\:  h(X,X) = 1\}
\]
Claim: $C > 0$. Suppose not. Then we can find a sequence $X_n \in T_{q_n}M$  with  $g(\nabla x^0, X_n) > 0$ and  $h(X_n, X_n) = 1$ such that $g(\nabla x^0, X_n) \to 0$. Since $U$ has compact closure, the $h$-unit bundle of $U$ within $TM$ has compact closure. Therefore there is a subsequence (still denoted by $X_n$) such that $X_n \to X_*$. Continuity implies $h(X_*,X_*) = 1$ and $g(\nabla x^0, X_*) = 0$. Since $\nabla x^0$ is past-directed,  $g(\nabla x^0, X_n) > 0$ implies each $X_n$ is future-directed causal. Hence $g(X_n, X_n) \leq 0$ and so continuity implies $g(X_*,X_*) \leq 0$. Therefore $X_*$ is either the zero vector or future-directed causal. Since $\nabla x^0$ is timelike, $g(\nabla x^0, X_*) = 0$ implies $X_*$ is either the zero vector or spacelike. Thus $X_*$ must be the zero vector, but this contradicts $h(X_*,X_*) = 1$. This proves the claim.

Fix $a < b$ in $I$ such that $\g(a), \g(b) \in U$. For $n$ sufficiently large, we have $\g_n(a), \g_n(b) \in U$. For these large $n$, we have
\[
x^0 \circ \g_n(b) - x^0 \circ \g_n(a) \,=\, \int_a^bg(\nabla x^0, \g_n') \,\geq\, C(b-a).
\]
By continuity we have $x^0 \circ \g(b) - x^0 \circ \g(a) \geq C(b-a)$. Therefore
\[
\frac{1}{b-a}\int_a^b g(\nabla x^0, \g') \,\geq\, C.
\] 
Since $a<b$ was arbitrary, Lebesgue's differentiation theorem implies $g(\nabla x^0, \g') \geq C > 0$ almost everywhere. Thus $\g' \neq 0$ almost everywhere for points in $I$ of $\gamma$ which lie in $U$.  Since $t_0 \in I$ was arbitrary, we have $\g' \neq 0$ almost everywhere in $I$. 
\qed

\medskip
\medskip

\begin{prop}\label{limit curve inextend prop}
Let $\g_n \colon \R \to M$ be a sequence of causal curves parameterized by $h$-arclength which converge to a causal curve $\g \colon \R \to M$ uniformly on compact subsets of $\R$. Then $\g$ is inextendible.
\end{prop}

\proof
Since $\g$ is a causal curve, it has an $h$-arclength reparameterization  $\tilde{\g} \colon (a,b) \to M$ by Proposition \ref{arclength prop}. Seeking a contradiction, suppose $\g$ is future extendible.  Then $\tilde{\g}$ is future extendible and so $b < \infty$ by Theorem \ref{inextend curves thm}.  Proposition \ref{endpoint prop} implies that there is a future endpoint $p \in M$ such that $\tilde{\g}$ extends continuously through $p$. By Proposition \ref{causal bounded by h prop}, there is an open set $U$ around $p$ such that $L_h(\lambda) < 1$ for all causal curves $\l \subset U$. Since $p$ is the future endpoint of $\g$, we have $\lim_{t \to \infty}\g(t) = p$ and hence there is some $t_0$ such that $\g\big([t_0,\infty)\big) \subset U$. Since the sequence $\g_n$ converges uniformly to $\g$ on compact subsets,  there exists an $N$ such that $\g_n\big([t_0,\, t_0+2]\big) \,\subset\,U$ for all $n \geq N$. The $h$-arclength of $\g_n|_{[t_0, t_0+2]}$ is $2$. But $2 > 1$. Therefore this contradicts our choice of open set $U$. Thus $\g$ is future inextendible. Likewise $\g$ is past inextendible.
\qed

\medskip
\medskip

\noindent\underline{\emph{Proof of Theorem \emph{\ref{limit curve thm}}}}:

\medskip

 Proposition \ref{existence limit curve prop} shows the existence of a locally Lipschitz curve $\g$. Proposition \ref{causal limit curve prop} shows that $\g$ is a causal curve. Proposition \ref{limit curve inextend prop} shows that $\g$ is inextendible.
  \qed

\medskip
\medskip

\section{Global hyperbolicity for $C^0$ spacetimes}

\subsection{Globally hyperbolic spacetimes}

\medskip

Fix a $C^0$ spacetime $(M,g)$. The \emph{Lorentzian length} of a causal curve $\g \colon I \to M$ is
 \[
 L(\g) \,=\, \int_I \sqrt{-g(\g', \g')}.
 \]
  If $\g$ is a causal curve from $p$ to $q$ such that its Lorentzian length satisfies $L(\g) \geq L(\l)$ for any other causal curve $\l$ from $p$ to $q$, then $\g$ is called a \emph{causal maximizer} from $p$ to $q$. In this section we will show that globally hyperbolic spacetimes always contain causal maximizers between causally related points (Theorem \ref{GH maximizer thm}). We first define global hyperbolicity.
  
\medskip
\medskip  
  
  \begin{Def}\label{globally hyperbolic def}
\emph{ $(M,g)$ is \emph{strongly causal at} $p$ if for every  neighborhood $U$ of $p$, there is a neighborhood $V \subset U$ of $p$ such that 
\[
\g(a), \g(b) \in V \:\:\:\: \Longrightarrow \:\:\:\: \g \subset U
\] 
whenever $\g \colon [a,b] \to M$ is a causal curve.  $(M,g)$ is \emph{strongly causal} if it is strongly causal at every point. If $(M,g)$ is strongly causal and  $J^+(p) \cap J^-(q)$ is compact for all $p,q \in M$, then $(M,g)$ is \emph{globally hyperbolic.}
}
\end{Def}

\medskip

\noindent\emph{Remarks.} 

\begin{itemize}

\item[-] A spacetime is \emph{causal} if there are no closed causal curves. A strongly causal spacetime implies that it is causal.

\item[-] For $C^2$ spacetimes strong causality can be weakened to causality  in the definition of global hyperbolicity \cite{BernalSanchez}, and if the spacetime dimension is greater than 2, then even causality is not needed \cite{HounnonkpeMinguzzi} . For $C^0$ spacetimes strong causality can be weakened to non-totally imprisoning \cite{Clemens}.

\end{itemize}

  \medskip
  \medskip

  The goal of this section is to prove the following fundamental result which was first established for continuous metrics by S{\"amann} in \cite{Clemens} and later proved independently in \cite{GLS}.

\medskip
\medskip

\begin{thm}\label{GH maximizer thm}
Let $(M,g)$ be a globally hyperbolic $C^0$ spacetime. Given any $q \in J^+(p)$ with $q \neq p$, there is a causal maximizer $\g$ from $p$ to $q$. Moreover $L(\g) < \infty$.
\end{thm}

\medskip
\medskip

\noindent\emph{Remarks.}
\begin{itemize}

\item[-] Causal maximizers in globally hyperbolic spacetimes have been used to establish spacetime inextendibility results in low regularity \cite{GLS, Sbierski2, GrafLing, MinguzziSuhr}.

\item[-] Theorem \ref{GH maximizer thm} shows that globally hyperbolic spacetimes are analogous to complete Riemannian manifolds: there is always a length-minimizing curve between any two points in a complete Riemannian manifold.
\end{itemize}

\medskip

\medskip

To prove Theorem \ref{GH maximizer thm}, we first establish some facts about strongly causal spacetimes. The following proposition shows that compact sets in a strongly causal spacetime cannot contain inextendible causal curves. This is often referred to as the `no imprisonment' property.

\medskip

\begin{prop}\label{imprisonment prop}
Suppose $(M,g)$ is strongly causal and $K \subset M$ is compact. Then
\[
\sup\, \{L_h(\g) \mid \g \subset K\} \,< \,\infty.
\]
\end{prop}

\proof
 By Proposition \ref{causal bounded by h prop}, for each $x \in K$, there is a neighborhood $U_x$ such that $L_h(\g) \leq 1$ for all $ \g \subset U_x$. By strong causality, there are  neighborhoods $V_x \subset U_x$ such that $\g \subset U_x$ whenever $\g \colon [a,b] \to M$ is a causal curve with endpoints in $V_x$.  Since $K$ is compact and covered by $\{V_x\}_{x \in K}$, there is a finite subcover $V_1, \dotsc, V_N$. 
 
Fix a causal curve $\g \colon [a,b] \to K$.   There exists some set from the finite cover which contains $\g(a)$, say $V_1$. Define $s_1$ via $s_1 = \sup \{ t \mid \g(t) \in V_1\}$. If $s_1 \neq b$, then $\g(s_1) \notin V_1$, and so there exists some set which contains $\g(s_1)$, say $V_2$. Define $s_2$ via $s_2 = \sup \{ t \mid \g(t) \in V_2\}$. If $s_2 \neq b$, then $\g(s_2) \notin V_2$, and so there exists some set which contains $\g(s_2)$, say $V_3$. And so on until $s^{k} = b \in V_k$ for some $1 \leq k \leq N$.  
  For $i = 1, \dotsc, k-1$, choose $t_1 < s_1$  such that $\g(t_i) \in V_{i} \cap V_{i+1}$. Define $\g_1 = \g|_{[a, t_1]}$ and $\g_2 = \g|_{[t_1, t_2]}$ and so on until $\g_k = \g|_{[t_{k-1}, b]}$. Then $\g_i \subset V_i$ and by the first paragraph of this proof, we have $L_h(\g_i) \leq 1$ for all $i = 1, \dotsc, k$. Therefore
  \[
   \pushQED{\qed}
  L_h(\g) \,=\, \sum_{i = 1}^k L_h(\g_i) \,\leq\,k \,\leq\,  N. \qedhere
\popQED
  \]

%
%
%

\medskip 
\medskip

The limit curve theorem guarantees the existence of a limit curve when there is \emph{one} accumulation point. The next proposition shows that for strongly causal spacetimes we can apply the limit curve theorem to \emph{two} accumulation points within a compact set.

\medskip
\medskip

\begin{prop}\label{limit curve for str cau prop}
Suppose $(M,g)$ is strongly causal and $K \subset M$ is a compact set. Let $\g_n \colon [0, b_n] \to K$ be a sequence of causal curves parameterized by $h$-arclength such that $\g_n(0) \to p$ and $\g_n(b_n) \to q$ with $q \neq p$. Then there is a $b \in (0, \infty)$ and a limit curve $\tilde{\g}\colon [0,b] \to M$ from $p$ to $q$ of $\tilde{\g}_n|_{[0,b]}$ where $\tilde{\g}_n \colon \R \to M$ are inextendible causal curve \linebreak extensions of $\g_n$.

\end{prop}

\proof
 By Theorem \ref{inextend curves thm}, we can extend $\g_n \colon [0, b_n] \to K$ to inextendible causal curves $\tilde{\g}_n \colon \R \to M$. Then $\tilde{\g}_n$ accumulates to $p$ at $0$. By the limit curve theorem, there is a subsequence (still denoted by $\tilde{\g}_n$) and a causal curve $\tilde{\g} \colon \R \to M$ witih $\tilde{\g}(0) = p$ such that $\tilde{\g}_n$ converges to $\g$ uniformly on compact subsets of $\R$. Since every sequence in $\R$ contains a monotone subsequence, we can assume $b_{n}$ is monotone by restricting to a further subsequence. Then either (1) $b_{n} \to \infty$ or (2) $b_{n} \to b < \infty$. The first scenario is ruled out by Proposition \ref{imprisonment prop}. Therefore the second scenario must hold. The triangle inequality gives
\begin{align*}
d_h\big(q, \tilde{\g}_{n}(b)\big) \,&\leq\, d_h\big(q, \g_{n}(b_{n})\big) \,+\, d_h\big(\g_{n}(b_{n}), \tilde{\g}_{n}(b)\big) 
\\
&\leq\, d_h\big(q, \g_{n}(b_{n})\big) \,+\, |b_n - b|.
\end{align*}
Since $\g_{n}(b_{n}) \to q$ and $b_{n} \to b$, the right hand side limits to 0. Thus $\tilde{\g}_{n}(b) \to q$. Since $q \neq p$, we have $b > 0$. Therefore $\tilde{\g}|_{[0,\, b]}$ is a causal curve from $p$ to $q$ which is a limit curve of $\tilde{\g}_n|_{[0,b]}$.
\qed

\medskip
\medskip

\begin{prop}
If $(M,g)$ is globally hyperbolic, then $J^+(p)$ is closed for all $p$. 
\end{prop}

\proof
Let $q$ be an accumulation point of $J^+(p)$. If $q = p$, then $q \in J^+(p)$ by definition. Suppose $q \neq p$. There is a sequence of points $q_n \to q$ and causal curves $\g_n$ from $p$ to $q_n$. Let $r \in I^+(q)$. Since $I^-(r)$ is open, there is an integer $N$ such that $n \geq N$ implies $q_n \in I^-(r)$. For these $n$, we have $\g_n \subset K$ where $K$ is the compact set $K = J^+(p) \cap J^-(r)$. Since $q \neq p$, Proposition \ref{limit curve for str cau prop} implies that there is a causal curve from $p$ to $q$.
\qed

\medskip
\medskip

The following technical proposition is needed for the proof of Theorem \ref{GH maximizer thm}.

\medskip
\medskip

\begin{prop}\label{causal bounded by g prop}
Given any $p \in M$ and $\e > 0$, there is a neighborhood $U$ such that $L(\g) < \e$  for all causal curves $\g \subset U$.
\end{prop}

\proof
The proof is similar to the proof of Proposition \ref{causal bounded by h prop}.
Fix $\e > 0$.  Choose a neighborhood $\phi \colon U_{3/5} \to \R^{n+1}$ from Lemma \ref{coord lem}. By continuity, we can shrink the neighborhood so that $|g_{\mu\nu}(x) - \eta_{\mu\nu}| < \e$ for all $x \in U_{3/5}$. Shrink $U_{3/5}$ even further so that $-\e < x^0 < \e$. Let $\g \subset U_{3/5}$ be any causal curve. Put $X = \g'$. Reparameterize $\g$ by $x^0$ so that $X^0 = 1$. Using $-g_{\mu\nu}(x) < -\eta_{\mu\nu} + \e$ and a similar calculation as in the proof of Lemma \ref{coord lem}, we have
\[
-g(X,X) \,<\, -\eta(X,X) + \e\sum_{\mu,\,\nu}|X^\mu X^\nu| \,<\, 1 + \e \left[1 + 2\sum_{i} |X^i| + \sum_{i,\,
j}|X^i X^j| \right].\]
Since $\eta^{-3/5}$ have `slope' 1/2, we have $|X^i| < 2$ for each $i$.
 Therefore  $-g(X,X) < 1 + \e C$ where $C = 1 + 4n + 4n^2$. Hence  $L(\g) = \int \sqrt{-g(\g', \g')}dx^0 < 2\e\sqrt{1 +\e C}$.
 \qed

\medskip
\medskip

The proof of Theorem \ref{GH maximizer thm} hinges on the upper semi-continuity of the Lorentzian length functional. This is Proposition \ref{length func prop}. We first use it to prove Theorem \ref{GH maximizer thm}. In section \ref{upper semi sec} we prove  Proposition \ref{length func prop}.

\medskip
\medskip

\begin{prop}[Upper semi-continuity of $L$]\label{length func prop}  Suppose $\g_n \colon [a,b] \to M$ is a sequence of causal curves which converge uniformly to a causal curve $\g \colon [a,b] \to M$. Given any $\e > 0$, there exists an integer $N$ such that $n \geq N$ implies $L(\g) \geq L(\g_n) - \e.$
\end{prop}

\medskip
\medskip

\noindent\underline{\emph{Proof of Theorem \emph{\ref{GH maximizer thm}}}}: 
 
 \medskip
 
Set
\[
\mathcal{L} = \sup \,\{\, L(\g)  \mid \g \text{ is a causal curve from } p \text{ to } q\}.\]
We first show $\mc{L}< \infty$. Let $K = J^+(p) \cap J^-(q)$. By Proposition \ref{causal bounded by g prop} for each $x \in K$, there is a neighborhood $U_x$ such that $L(\g) \leq 1$ for all $ \g \subset U_x$. By strong causality, there are  neighborhoods $V_x \subset U_x$ such that $\g \subset U_x$ whenever $\g \colon [a,b] \to M$ is a causal curve with endpoints in $V_x$.   Since $K$ is compact and covered by $\{V_x\}_{x \in K}$, there is a finite subcover $V_1, \dotsc, V_N$. Therefore $\mc{L}$ is bounded by $N$ via the same proof used in Proposition \ref{imprisonment prop}. 

By definition of $\mc{L}$ there is a sequence of causal curves $\g_n\colon [0, b_n] \to M$ from $p$ to $q$ \linebreak satisfying $\mc{L} \leq L(\g_n) + 1/n.$ Let $\g_n$ be parameterized by $h$-arclength. Proposition \ref{limit curve for str cau prop} shows that there is a $b \in (0,\infty)$ and a limit curve $\tilde{\g}\colon [0,b] \to M$ from $p$ to $q$ of $\tilde{\g}_n|_{[0,b]}$ where $\tilde{\g}_n \colon \R \to M$ are inextendible causal curve extensions of $\g_n$. By restricting to a subsequence, we can assume $\tilde{\g}_n|_{[0,b]}$ converges uniformly to $\tilde{\g}$. By upper semi-continuity of the length functional, given any $\e > 0$ there exists an $N$ such that $n \geq N$ implies 
\begin{align*}
L(\tilde{\g}) + \e \,&\geq \, L(\tilde{\g}_n|_{[0,b]}) 
\\
&=\, L(\g_n) \,+\, \int_{b_n}^{b}\sqrt{-g(\tilde{\g}_n',\tilde{\g}_n')} \\
&\geq \, (\mc{L} - 1/n) \,+\, \int_{b_n}^{b}\sqrt{-g(\tilde{\g}_n',\tilde{\g}_n')} .
\end{align*}
Since this is true for all $n \geq N$, we have $L(\tilde{\g}) + \e \geq \mc{L}$ (note we used Proposition \ref{locally lipschitz components prop} here). Since $\e$ was arbitrary, we have $L(\tilde{\g}) \geq \mathcal{L}$. Thus $\tilde{\g}$ is a causal maximizer from $p$ to $q$.
\qed

\medskip
\medskip

\subsection{Cauchy surfaces imply global hyperbolicity}

\medskip
Fix a $C^0$ spacetime $(M,g)$ with a complete Riemannian metric $h$. In this section we show that global hyperbolicity is implied by the more familiar notion of a Cauchy surface:

\medskip
\medskip

\begin{thm}\label{CS implies GH thm}
Let $(M,g)$ be a $C^0$ spacetime. If $(M,g)$ has a Cauchy surface, then $(M,g)$ is globally hyperbolic.
\end{thm}

\medskip
\medskip

\noindent\emph{Remark}. In fact global hyperbolicity is equivalent to the existence of a Cauchy surface even for continuous metrics \cite{Clemens}. 

\medskip
\medskip

\begin{Def}
\emph{A \emph{Cauchy surface} for $(M,g)$ is a set $S \subset M$ such that every inextendible causal curve intersects $S$ exactly once.}
\end{Def}

\medskip
\medskip

Recall that a subset $S \subset M$ is \emph{achronal} if $I^+(S) \cap S = \emptyset$. The \emph{edge} of an achronal set $S$ is the set of points $p \in \ov{S}$ such that for every neighborhood $U$ of $p$, there is a timelike curve $\g \colon [a,b] \to U$ such that $\g(a) \in I^-(p,U)$, $\g(b) \in I^+(p,U)$, and $\g \cap S = \emptyset$.

\medskip
\medskip

\begin{prop}
If $S$ is a Cauchy surface, then $S$ is achronal and has empty edge.
\end{prop}

\proof
We first show $S$ is achronal. If $I^+(S) \cap S \neq \emptyset$, then there would be a timelike curve $\g \colon [a,b] \to M$ with endpoints on $S$. We can extend $\g$ to an inextendible causal curve $\tilde{\g}$ via Theorem \ref{inextend curves thm}. But then $\tilde{\g}$ intersects $S$ twice which contradicts the definition of a Cauchy surface. Therefore $S$ is achronal.

Now we show $S$ has no edge points. Seeking a contradiction, suppose $p \in \ov{S}$ is an edge point of $S$. Let $U$ be a neighborhood of $p$. Then there is a timelike curve $\g \colon [a,b] \to U$ such that $\g(a) \in I^-(p, \, U)$, $\g(b) \in I^+(p, \,U)$, and $\g \cap S = \emptyset$. We can extend $\g$ to an inextendible causal curve $\tilde{\g} \colon \R \to M$. Since $S$ is a Cauchy surface, there exists some $t_0$ such that $\tilde{\g}(t_0) \in S$. By assumption we have $t_0 \notin [a,b]$. Suppose $t_0 < a$. Since $p \in \ov{S}$, there is a sequence of points $p_n \in S$ such that $p_n \to p$. Therefore for all sufficiently large $n$, we have $p_n \in I^+\big(\g(a), U\big)$. Then there is a causal curve from $\tilde{\g}(t_0)$ to $\g(a)$ to $p_n$. This contradicts the definition of a Cauchy surface. Likewise supposing $t_0 > b$ yields a contradiction.
\qed

\medskip
\medskip

\begin{cor}\label{Cauchy surface is top sur}
If $S$ is a Cauchy surface, then $S$ is a $C^0$ hypersurface.
\end{cor}

\proof
This follows from Theorem \ref{achronal edge thm}.
\qed

\medskip
\medskip

\begin{lem}\label{CS decomp lem} Let $S$ be a Cauchy surface. Then $S$ separates $M$ via the disjoint union
\[M = I^+(S) \,\sqcup  \, S \, \sqcup\, I^-(S).
\]
\end{lem}

\proof
Fix $p \in M$.  Since $M$ is time-oriented, there is a $C^1$ timelike vector field $X$ on $M$. Let $\g_p$ denote the maximal integral curve of $X$ through $p = \g_p(0)$. Since $\g_p$ is maximal and hence inextendible as a continuous curve, it is an inextendible causal curve. Therefore it must intersect $S$ at some point $t_0$. If $t_0 = 0$, then $p \in S$. If $t_0 > 0$, then $\g_p$ is $C^1$ on $[0,t_0]$ and hence $\g_p|_{[0,t_0]}$ is a timelike curve by Proposition \ref{normal causal and timelike prop}. Thus, if $t_0 > 0$, then $p \in I^-(S)$. Likewise, if $t_0 < 0$, then $p \in I^+(S)$. The disjointness follows from $S$ being achronal. 
\qed

\medskip
\medskip

\begin{cor}
If $S$ is a Cauchy surface, then $S$ is closed. 
\end{cor}

\medskip
\medskip

\begin{prop}\label{inext curves intersect prop}
Let $S$ be a Cauchy surface. Then every inextendible causal curve \linebreak intersects $I^+(S)$ and $I^-(S)$.
\end{prop}

\proof
Let $\g\colon \R \to M$ be an inextendible causal curve. Seeking a contradiction, suppose $\g$ does not intersect $I^+(S)$. Since $S$ is a Cauchy surface, there exists some $t_0$ such that $\g(t_0) \in S$. Let $t_1 > t_0$. By Lemma \ref{CS decomp lem}, we have $\g(t_1) \in S \cup I^-(S)$. If $\g(t_1) \in S$, then there is a causal curve from $\g(t_0)$ to $\g(t_1)$ which is a contradiction. If $\g(t_1) \in I^-(S)$, then there is a causal from $\g(t_0)$ to $\g(t_1)$ to a point on $S$ -- again a contradiction. Therefore $\g$ must intersect $I^+(S)$. Likewise $\g$ must intersect $I^-(S)$.
\qed

\medskip
\medskip

\begin{prop}\label{future inextend intersect prop}
Let $S$ be a Cauchy surface. Let $\g \colon [0,\infty) \to M$ be a future inextendible causal curve. Then $\g$ intersects $I^+(S)$.
\end{prop}

\proof
Extend $\g$ to an inextendible causal curve $\tilde{\g} \colon \R \to M$ via Theorem \ref{inextend curves thm}. If the conclusion did not hold, then Proposition \ref{inext curves intersect prop} implies there exists a $t_0 < 0$ such that $\tilde{\g}(t_0) \in I^+(S)$. By Lemma \ref{CS decomp lem}, we must have $\g(0) \in S \cup I^-(S)$. But this implies we can find a causal curve which intersects $S$ twice.
\qed

\medskip
\medskip

\begin{lem}\label{Cauchy h-length bound lem}
Let $S$ be a Cauchy surface. If $p \in I^+(S)$, then
\[
\sup\,\{ L_h(\g) \mid \g \text{ is a causal curve from } S \text{ to } p\} \,< \,\infty.
\]
\end{lem}

\proof
Suppose this is not true. Then we can find a sequence of $h$-arclength parameterized causal curves $\g_n \colon [a_n, 0] \to M$ from $S$ to $p$ such that $a_n \to - \infty$. By Theorem \ref{inextend curves thm} we can extend these curves to inextendible causal curves $\tilde{\g}_n \colon \R \to M$. By the limit curve theorem there is a subsequence (still denoted by $\tilde{\g}_n$)  which converges to an inextendible causal curve $\g \colon \R \to M$ passing through $p = \g(0)$. 

We will show $\g \subset S \cup I^+(S)$. Consider $t_0 \geq 0$. Since $p \in I^+(S)$, we have $\g(t_0) \notin I^-(S)$. Therefore  $\g(t_0) \in S \cup I^+(S)$ by Lemma  \ref{CS decomp lem}. Now consider $t_0 < 0$. Seeking a contradiction, suppose $\g(t_0) \in I^-(S)$. Since $I^-(S)$ is open, $\tilde{\g}_n(t_0) \in I^-(S)$ for all sufficiently large $n$.  Since $a_n \to - \infty$, we can choose $n$ large enough so that $a_n < t_0$. Then $\tilde{\g}_n(t_0) = \g_n(t_0)$. Therefore we have a causal curve from $\g_n(a_n)$ to $\g_n(t_0) \in I^-(S)$ which contradicts the definition of a Cauchy surface. Thus we have shown $\g \subset S \cup I^+(S)$, but this contradicts Proposition \ref{inext curves intersect prop}.
\qed

%

\medskip
\medskip

\begin{lem}\label{no closed causal curves lem}
Let $S$ be a Cauchy surface. Then there are no closed causal curves in $M$.
\end{lem}

\proof
Let $\g \colon [0,b] \to M$ be a closed causal curve. We define an inextendible causal curve $\tilde{\g} \colon \R \to M$ by $\tilde{\g}|_{[0,b]} = \g$ and $\tilde{\g}(t + b) = \tilde{\g}(t)$ for all $t$. Since $\tilde{\g}$ is inextendible it must intersect $S$. Since $\g$ is closed, it intersects $S$ infinitely often. This contradicts the definition of a Cauchy surface. 
\qed

\medskip
\medskip

\medskip
\medskip

\begin{prop}\label{CS strong caus prop}
Let $S$ be a Cauchy surface. Then $(M,g)$ is strongly causal.
\end{prop}

\proof
Suppose strong causality failed at the point $p$. Then there is a neighborhood $U$ and a sequence of causal curves $\g_n\colon [0,b_n] \to M$ parameterized by $h$-arclength such that $\g_n(0) \to p$ and $\g_n(b_n) \to p$ but each $\g_n$ leaves $U$. Note this implies there is a $c > 0$ such that $b_n > c$ for all $n$. Extend each $\g_n$ to inextendible causal curves $\tilde{\g}_n \colon \R \to M$ via Theorem \ref{inextend curves thm}. Since $\g_n(0) \to p$, the limit curve theorem yields a limit curve $\g \colon \R \to M$ of the $\tilde{\g}_n$ such that $\g(0) = p$. Therefore there is a subsequence (still denoted by $\tilde{\g}_n$) such that $\tilde{\g}_n$ converges to $\g$ uniformly on compact subsets. By restricting to a further subsequence, we either have (1) $b_n \to \infty$ or (2) $b_n \to b < \infty$. Suppose the second case. Then the triangle inequality gives
\begin{align*}
d_h\big(\g(b), p\big) \,&\leq\, d_h\big(\g(b), \tilde{\g}_n(b)\big) \,+\, d_h\big(\tilde{\g}_n(b), \tilde{\g}_n(b_n)\big) \,+\, d_h\big(\tilde{\g}_n(b_n), p\big).
\end{align*}
Each of the terms on the right hand side limits to $0$. Therefore $\g(b) = p$. Since $b \geq c > 0$, we have a closed causal curve through $p$. This contradicts Lemma \ref{no closed causal curves lem}.

Therefore we must have $b_n \to \infty$. Proposition \ref{future inextend intersect prop} implies that there exists a $t_0 \geq 0$ such that $\g(t_0) \in I^+(S)$. By passing to a further subsequence, we can assume $b_n \geq t_0$. Therefore $\tilde{\g}_n(t_0) = \g_n(t_0)$. Fix $q \in I^+(p)$. There exists an $N$ such that $n \geq N $ implies $\g_n(b_n) \in I^-(q)$ and $\g_n(t_0) \in I^+(S)$ since these are open sets. Therefore for these $n$, there is a causal curve $\l_n$ from $S$ to $\g_n(t_0)$ to $\g_n(b_n)$ to $q$. But $L_h(\l_n) \to \infty$ since $b_n \to \infty$. This contradicts Lemma \ref{Cauchy h-length bound lem} provided we show $q \in I^+(S)$. Indeed since $q \in J^+(S)$, we have $q \notin I^-(S)$. Therefore $q \in S \cup I^+(S)$ by Lemma \ref{CS decomp lem}. But $q \notin S$ otherwise $\l_n$ would be a causal curve which intersects $S$ twice. Hence $q \in I^+(S)$.
\qed

\medskip
\medskip

\begin{lem}\label{diamond bounded pre lem}
Let $S$ be a Cauchy surface. Then for all $p \in I^-(S)$ and $q \in I^+(S)$, we have 
\[
\sup\,\{ L_h(\g) \mid \g \text{ is a causal curve from } p \text{ to } q\} \,< \,\infty.
\]
\end{lem}

\proof
By Lemma \ref{Cauchy h-length bound lem}, we have 
$\sup\{ L_h(\g) \mid \g \text{ is causal from } S \text{ to } q\} = b < \infty$. Likewise the time-dual of Lemma \ref{Cauchy h-length bound lem} gives $\sup\{ L_h(\g) \mid \g \text{ is causal from } p \text{ to } S\} = a < \infty$. Then any causal curve from $p$ to $q$ has $h$-arclength bounded by $a + b$. 
\qed

\medskip
\medskip

\begin{lem}\label{diamond bounded lem}
Let $S$ be a Cauchy surface. Then for all $p$ and $q$ we have
\[
\sup\,\{ L_h(\g) \mid \g \text{ is a causal curve from } p \text{ to } q\} \,< \,\infty.
\]
\end{lem}

\proof
Seeking a contradiction, suppose the supremum was infinite. By Proposition \ref{future inextend intersect prop}, there is a causal curve from $q$ to $q' \in I^+(S)$. Likewise the time-dual of Proposition \ref{future inextend intersect prop} guarantees a point $p' \in I^-(S)$ and a causal curve from $p'$ to $p$. Since the supremum is infinite, we have $\sup\{ L_h(\g) \mid \g \text{ is causal from } p' \text{ to } q'\} = \infty$. But this contradicts Lemma \ref{diamond bounded pre lem}.
\qed

\medskip
\medskip

\begin{prop}\label{J^+ is closed prop}
Let $S$ be a Cauchy surface. Then $J^+(p)$ is closed for all $p\in M$.
\end{prop}

\proof
Let $q$ be an accumulation point of $J^+(p)$. If $q = p$, then $q \in J^+(p)$ by definition. Assume $q \neq p$. Then there is a sequence $h$-arclength parameterized causal curves $\g_n \colon [0, b_n] \to M$ from $p$ to $q_n$ where $q_n \to q$. Extend these to inextendible causal curves $\tilde{\g}_n \colon \R \to M$ via Theorem \ref{inextend curves thm}. By the limit curve theorem, there is a subsequence (still denoted by $\tilde{\g}_n$) which converges to an inextendible causal curve $\g \colon \R \to M$ with $\g(0) = p$. By restricting to a further subsequence, we either have \linebreak (1) $b_n \to \infty$ or (2) $b_n \to b < \infty$. It suffices to show that only (2) can hold. For then the same triangle inequality argument used in the proof of Proposition \ref{CS strong caus prop} implies $\g \colon [0,b] \to M$ is a causal curve from $p$ to $q$.

Seeking a contradiction, suppose (1) holds. Let  $q_0 \in I^+(q)$. Then for all sufficiently large $n$, we have $q_n \in I^-(q_0)$. Therefore there are causal curves $\l_n$ from $p$ to $q_0$ with $L_h(\l_n) \to \infty$. But this contradicts Lemma \ref{diamond bounded lem}.
\qed

\medskip
\medskip

\newpage

\noindent\underline{\emph{Proof of Theorem \emph{\ref{CS implies GH thm}}}}:

\medskip

By Proposition \ref{CS strong caus prop} we have $(M,g)$ is strongly causal. By Proposition \ref{J^+ is closed prop} we have  $J^+(p) \cap J^-(q)$ is closed for all $p$ and $q$. So by the Hopf-Rinow theorem, it suffices to show that this intersection is bounded with respect to $d_h$. This follows from Lemma \ref{diamond bounded lem}.
\qed

\medskip
\medskip

\subsection{Upper semi-continuity of the Lorentzian length functional}\label{upper semi sec}

This section is solely devoted to proving the upper semi-continuity of the Lorentzian length functional. This is Proposition \ref{length func prop} which played a chief role in the proof of Theorem \ref{GH maximizer thm}. 
\medskip
\medskip

\noindent\emph{Remark.} The proof of upper semi-continuity in \cite{GLS} used approximating smooth metrics and the fact that the Lorentzian length of a causal curve can be found by taking the length of a limit of interpolating geodesics. This last fact was somewhat of a folklore theorem until Minguzzi proved it in \cite[Theorem 2.37]{MinguzziLivRev}. The proof in this section relies on similar ideas but instead of approximating via geodesics, we approximate via $\eta^{-\e}$-maximizers (see Definition \ref{interpolate def}).

\medskip
\medskip

Fix a $C^0$ spacetime $(M,g)$. Let $\phi \colon U_\e \to \R^{n+1}$ be a coordinate system as in Lemma \ref{coord lem}, and $B_\e \subset U_\e$ denote an open set satisfying the same properties (1) - (5) as in Lemma \ref{coord lem}, but in addition $\phi(B_\e)$ is also a Euclidean ball with respect to the usual norm on $\R^{n+1}$ and also $B_\e$ has compact closure. $L$ will denote the Lorentzian length with respect to $g$ while $L_{\eta^{-\e}}$ will denote the Lorentzian length with respect to $\eta^{-\e}$. Given two points $p,q \in B_\e$, the \emph{straight line} joining $p$ to $q$ is the unique curve $\lambda \colon [0,1] \to B_\e$ such that $\phi \circ \lambda(t) = t\phi(q) + (1-t)\phi(p)$. 

\medskip
\medskip

\begin{prop}\label{eta-maximizer prop}
Suppose $\g \colon [a,b] \to B_\e$ is a causal curve. Let $\lambda \subset B_\e$ be the straight line joining $\g(a)$ to $\g(b)$. Then
\begin{itemize}

\item[\emph{(1)}] $\lambda$ is an $\eta^{-\e}$-timelike curve,

\item[\emph{(2)}] $L_{\eta^{-\e}}(\lambda) \,\geq\, L_{\eta^{-\e}}(\g)$.

\end{itemize}

\end{prop}

\proof
We first prove (1). Write $\tilde{\g} = \phi \circ \g$ and $\tilde{B}_\e = \phi \circ B_\e$. Since $\g$ is a causal curve, it is an $\eta^{-\e}$-timelike curve by Lemma \ref{coord lem}. Therefore, by Lemma \ref{euclidean cone lem}, we have $\tilde{\g}(b) \in C^+_{-\e'}\big(\tilde{\g}(a), \tilde{B}_\e\big)$ for all $\e' \in (\e, 1)$. Therefore $\tilde{\g}(b)$ lies in the closure $\ov{C}$ where $C = C^+_{-\e}\big(\tilde{\g}(a), \tilde{B}_\e\big)$. If $\tilde{\g}(b)$ lies within the interior of $C$, then $\l$ is $\eta^{-\e}$-timelike. If $\tilde{\g}(b)$ lies on the boundary $\pd C$, then $\l$ is $\eta^{-\e}$-null. Therefore it suffices to show $\tilde{\g}(b)$ must lie within the interior. Seeking a contradiction, suppose $\tilde{\g}(b) \in \pd C$. Let $U \subset I^+_{\eta^{-\e}}\big(\g(a), B_\e\big)$ be an open set around $\g(b)$. Since $\tilde{\g}(b)$ lies on the boundary $\pd C$, there exists a point $\tilde{p} \in \tilde{U} \setminus \ov{C}$ where $\tilde{U} = \phi \circ U$. But since $p = \phi^{-1}(\tilde{p}) \in I^+_{\eta^{-\e}}\big(\g(a), B_\e\big)$, Lemma \ref{euclidean cone lem} implies $\tilde{p} \in C^+_{-\e'}\big(\tilde{\g}(a), \tilde{B}_\e\big)$ for all $\e' \in (\e, 1)$. Hence $\tilde{p} \in \ov{C}$ which is a contradiction.

Now we prove (2). From (1) we know that $\tilde{\l} = \phi \circ \l$ is an $\eta^{-\e}$-timelike curve which is the straight line joining $\tilde{\g}(a)$ to $\tilde{\g}(b)$. Since $\tilde{\l}$ is $\eta^{-\e}$-timelike, there is an $\eta^{-\e}$-isometry $\psi \colon \R^{n+1} \to \R^{n+1}$ which  sends $\tilde{\g}(a)$ to the origin and takes $\tilde{\g}(b)$ to $\psi \circ \tilde{\g}(b)$ which lies on the positive $x^0$-axis. Note that $\psi$ is composed of a translation and an $\eta^{-\e}$-Lorentz transformation. Similarly we have $\psi \circ \tilde{\l}$ lies entirely on the positive $x^0$-axis. Parameterize $\tilde{\g}$ and $\tilde{\l}$ by $x^0$ and put $X = (\psi \circ \tilde{\g})'$. Since $\psi$ is an $\eta^{-\e}$-isometry, we have
\[
\pushQED{\qed}
L_{\eta^{-\e}}(\g) \,=\, \int \sqrt{\frac{1+\e}{1-\e} - \delta_{ij}X^iX^j} \,\leq\,  \int \sqrt{\frac{1+\e}{1-\e}} \,=\, L_{\eta^{-\e}}(\l).   \qedhere
\popQED
\]

\medskip
\medskip

\noindent Proposition \ref{eta-maximizer prop} justifies the terminology in the next definition.

\medskip
\medskip

\begin{Def}\label{interpolate def} \emph{Let $\g \colon [a,b] \to B_\e$ be a causal curve.}
\:
\begin{itemize}
\item[-] \emph{Let $P$ be a partition $a = t_0 < t_1 < \dotsb < t_k = b$ of the interval $[a,b]$. The \emph{interpolating $\eta^{-\e}$-maximizer} of $\g$ with respect to $P$ is the $\eta^{-\e}$-causal curve $\l \subset B_\e$ formed by concatenating the straight lines joining $\g(t_{i-1})$ to $\g(t_i)$ for $i =  1, \dotsc, k$.
} 

\item[-] \emph{For $k = 1, 2, \dotsc$ let $P_k$ denote the partition $a = t_0 < t_1 < \dotsc< t_{k} = b$ such that $t_{i} - t_{i-1} = (b-a)/k$ for all $i = 1, \dotsc, k$. Let $\lambda_k$ denote the interpolating $\eta^{-\e}$-maximizer of $\gamma$ with respect to $P_k$.}
\end{itemize}
\end{Def}

\medskip
\medskip

\noindent The following proof is inspired by \cite[Theorem 2.37]{MinguzziLivRev}.

\medskip
\medskip

\begin{prop}\label{approx length prop}
Let $\g \colon [a,b] \to B_\e$ be a causal curve. Then 
\[
L_{\eta^{-\e}}(\g) \,=\, \lim_{k\to \infty}L_{\eta^{-\e}}(\l_k).
\]
\end{prop}

\proof
The proof is an application of Lebesgue's dominated convergence theorem.  Assume $\g$ and $\l_k$ are parameterized by the time function $x^0$ on $B_\e$. Then $-\eta^{-\e}(\l_k', \l_k') \leq (1+\e)/(1-\e)$. Thus it suffices to show $\l_k' \to \g'$ almost everywhere.
 Let $A$ denote the set of points in $[a,b]$ where $\g$ is not differentiable. Likewise, let $A_k$ denote the set of points in $[a,b]$ where $\l_k$ is not differentiable. Then $D = [a,b] \setminus \left(A \cup \bigcup_k A_k \right)$ has full measure and represents the set of differentiable points which belong to $\g$ and each $\l_k$.

Fix $t_* \in D$. We will show $\l_k'(t_*) \to \g'(t_*)$. Let $\tilde{\g} = \phi \circ \g$ and $\tilde{\l}_k = \phi \circ \l_k$ where $\phi$ is the coordinate map for $B_\e$.
Let $a_k$ be the greatest point on the partition $P_k$ such that $a_k < t_*$ and $b_k$ be the least point on the partition such that $t_* < b_k$. Then $\g(a_k) = \l_k(a_k)$ and $\g(b_k) = \l_k(b_k)$. Therefore the triangle inequality gives
\begin{align*}
\big|\tilde{\g}'(t_*) - \tilde{\l}_k'(t_*)\big| \,&\leq\, \left|\tilde{\g}'(t_*) - \frac{\tilde{\g}(b_k) - \tilde{\g}(a_k)}{b_k - a_k}\right| \,+\, \left| \frac{\tilde{\l}_k(b_k) - \tilde{\l}_k(a_k)}{b_k - a_k} - \tilde{\l}_k'(t_*)\right|
\\
&=\,\left|\tilde{\g}'(t_*) - \frac{\tilde{\g}(b_k) - \tilde{\g}(a_k)}{b_k - a_k}\right| \,+\, 0.
\end{align*}
We get zero for the second term on the right hand side since $\tilde{\l}_k$ is composed of straight lines. Now we use the definition of the derivative to bound the first term on the right hand side. First note the identity
\begin{align*}
&\tilde{\g}'(t_*) - \frac{\tilde{\g}(b_k) - \tilde{\g}(a_k)}{b_k - a_k} 
\\
\,=\, &\frac{b_k -t_*}{b_k - a_k}\left(\tilde{\g}'(t_*)-\frac{\tilde{\g}(b_k) - \tilde{\g}(t_*)}{b_k - t_*} \right) + \frac{t_* - a_k}{b_k - a_k}\left(\tilde{\g}'(t_*)-\frac{\tilde{\g}(t_*) - \tilde{\g}(a_k)}{t_*-a_k} \right).
\end{align*}
Fix $\e_0 > 0$. By definition of the derivative, there exists a $\delta > 0$ such that $|t- t_*| < \delta$ implies
\[
\left|\tilde{\g}'(t_*) - \frac{\tilde{\g}(t_*) - \tilde{\g}(t)}{t_* - t}\right| \,<\, \e_0 .
\]
Choose $N$ large enough so that $k \geq N$ implies $t_* - a_k < \delta$ and $b_k - t_* < \delta$. Then using the identity above, we have
\[
\big|\tilde{\g}'(t_*) - \tilde{\l}_k'(t_*)\big| \,\leq\,
\left|\tilde{\g}'(t_*) - \frac{\tilde{\g}(b_k) - \tilde{\g}(a_k)}{b_k - a_k}\right| \,\leq\, \frac{b_k - t_*}{b_k-a_k}\e_0 + \frac{t_* - a_k}{b_k-a_k}\e_0 \,=\, \e_0.
\]
Thus $\l_k'(t_*) \to \g'(t_*)$ as desired.
\qed

\medskip
\medskip

Lemma \ref{L and L_h lem} will be used to prove the upper semi-continuity of $L$ locally (Lemma \ref{lem for upper semi prop}).   Then Lemma \ref{lem for upper semi prop} will be used to prove the upper semi-continuity of $L$ globally.

\medskip
\medskip

\begin{lem}\label{L and L_h lem}
Given any $p \in M$, there is a neighborhood $U$ around $p$ and a constant  $C$  such that for any $p_0 \in U$ and any $0 < \e < 3/5$, there is a neighborhood $B_\e \subset U$ centered around $p_0$ such that
$\big|L(\g) - L_{\eta^{-\e}}(\g)\big| \leq CL_h(\g)\sqrt{\e}$ for any causal curve $\g \subset B_\e$.
\end{lem}

\proof  Fix $p \in M$. Let $U_{3/5}$ be a neighborhood from Lemma \ref{coord lem} with compact closure. Let $(x^0, x^1, \dotsc, x^n)$ denote the coordinates for $U_{3/5}$. From (3) of Lemma \ref{coord lem}, we have $|g_{00}(x) - \eta_{00}| < 3/5$ for all $x \in U_{3/5}$. Hence $|g_{00}(x)| < 8/5$. 

Define $c$ via $1/c^2 \,=\, \inf \,\{h(X,X) \mid x \in U_{3/5},\: X \in T_xM,\: \delta_{\mu\nu}X^\mu X^\nu = 1\}.$
Since $U_{3/5}$ has compact closure, we have $1/c^2 > 0$. Then for all $x \in U_{3/5}$ and $X \in T_xM$, we have $
\delta_{\mu\nu}X^\mu X^\nu \,\leq\, c^2h(X,X).$
In particular, if $h(X,X) = 1$, then $|X^0|^2 \leq \delta_{\mu\nu}X^\mu X^\nu \leq c^2$.

Fix $p_0 \in U_{3/5}$ and $0< \e < 3/5$. By Lemma \ref{coord lem}, there is a $B_\e \subset U_{3/5}$ neighborhood around $p_0$ with coordinates $(y^0, y^1, \dotsc, y^n)$, but we apply the Gram-Schmidt process  so that $\pd / \pd y^0$ is parallel to $\pd /\pd x^0$ at $p_0$. Then $x^0 =  y^0/\sqrt{|g_{00}(p_0)|}$ and $\pd x^i/ \pd y^0 = 0$ on $B_\e$. Let $\tilde{g}_{\mu\nu} = g(\frac{\pd}{\pd y^\mu}, \frac{\pd}{\pd y^\nu})$ denote the components of $g$ with respect to $(y^0, y^1, \dotsc, y^n)$ to distinguish them from $g_{\mu\nu}$. Then by construction, we have $\tilde{g}_{00} = -1$ at $p_0$.

Let $\g \colon [0,b] \to B_\e$ be any causal curve parameterized by $h$-arclength. Let $X = \g'$. In components we write $X = X^\mu \frac{\pd}{\pd x^\mu} = \tilde{X}^\mu \frac{\pd}{\pd y^\mu}$. Then $|\tilde{g}_{\mu\nu}(x) - \eta_{\mu\nu}| < \e$ for all $x \in B_\e$ implies $|g(X,X) - \eta(X,X)| < \e \sum_{\mu, \nu}|\tilde{X}^\mu \tilde{X}^\nu|$. Then using $\eta = \eta^{-\e} + \frac{2\e}{1-\e}(dy^0)^2$, we have
\begin{align*}
-\eta^{-\e}(X,X) \,&<\, -g(X,X) \,+\, \frac{2\e}{1-\e}|\tilde{X}^0|^2 \,+\, \e\sum_{\mu,\nu}|\tilde{X}^\mu \tilde{X}^\nu|
\\
&=\, -g(X,X) \,+\, \frac{2\e}{1-\e}|\tilde{X}^0|^2 \,+\, \e|\tilde{X}^0|^2 \left[1 + \sum_i \frac{|\tilde{X}^i|}{|\tilde{X}^0|} + \sum_{i,j} \frac{|\tilde{X}^i \tilde{X}^j|}{|\tilde{X}^0|^2}  \right]
\end{align*}
Since $0 < \e < 3/5$, we have $2/(1-\e) < 5$. Also, the term in the bracket is less than $1 + 4n + 4n^2$ where $n+1$ is the dimension of the spacetime. Therefore 
\[
-\eta^{-\e}(X,X) \,<\, -g(X,X) + \e |\tilde{X}^0|^2\left(6 + 4n + 4n^2 \right).
\]
By the first and second paragraphs of this proof, we have
\[
\tilde{X}^0(t) \,=\, (y^0 \circ \g)'(t) \,=\, (x^0 \circ \g)'(t)\sqrt{|g_{00}(p_0)|} \,<\, c\sqrt{8/5}.
\]
Thus
\[
- \eta^{-\e}(X,X) \,<\, -g(X,X) \,+\, \e C^2
\]
where $C^2 = c^2(8/5)(6 + 4n + 4n^2)$. This establishes $L_{\eta^{-\e}}(\g) < L(\g) + CL_h(\g) \sqrt{\e}$. The proof of the other inequality is analogous.
\qed

\medskip
\medskip

\begin{lem}\label{lem for upper semi prop}
Given any $p \in M$, there is a neighborhood $U$ around $p$ such that given any $\e > 0$ and any sequence of causal curves $\g_n \colon [a,b] \to U$  which converge uniformly to the causal curve $\g \colon [a,b] \to U$, there exists an $N$ such that $n \geq N$ implies $L(\g) \geq L(\g_n) - \e$. 
\end{lem}

\proof
Fix $p \in M$ and let $U$ be the neighborhood from Lemma \ref{L and L_h lem}. Let $\g_n \colon [a,b] \to U$ be a sequence of causal curves parameterized by $h$-arclength which converge uniformly to a causal curve $\g \colon [a,b] \to U$.

 Fix $\e > 0$. For any $p_0 \in \g\big([a,b]\big)$, there is a $B_\e \subset U$ around $p_0$ from Lemma \ref{L and L_h lem}. Since $\g\big([a,b]\big)$ is compact, we can cover it by finitely many neighborhoods $B^1_\e \dotsc, B^l_\e$. We order these sets in the following way: There exists some set which contains $\g(a)$, say $B^1_\e$. Define $s^1$ via $s^1 = \sup \{ t \mid \g\big([a,t)\big) \subset B^1_\e\}$. Either $s^1 = b$ in which case we take $l = 1$ and stop. Otherwise $\g(s^1) \notin B^1_\e$, and so there exists some set which contains $\g(s^1)$, say $B^2_\e$. Define $s^2$ via $s^2 = \sup \{ t \mid \g\big([s^1,t)\big) \subset B^2_\e\}$. Either $s^2 = b$ in which case we take $l = 2$ and stop. Otherwise $\g(s^2) \notin B^2_\e$, and so there exists some set which contains $\g(s^2)$, say $B^3_\e$. And so on. Repeat this process until $s^{l} = b \in B^l_\e$.  For $i = 1, \dotsc, l-1$, choose $t^i < s^i$  such that $\g(t^i) \in B^{i}_\e \cap B^{i+1}_\e$. Define $\g^1 = \g|_{[a, t^1]}$ and $\g^2 = \g|_{[t^1, t^2]}$ and so on until $\g^l = \g|_{[t^{l-1}, b]}$. Then $\g^i \subset B^i_\e$ for all $i = 1, \dotsc, l$. 

Let $t^0 = a$ and $t^l = b$.  For each $i = 1, \dotsc, l$, we partition the domain $[t^{i-1}, t^i]$ into $k$ equal subintervals (i.e. via $P_k$) and let $\l^i_k$ denote the interpolating $\eta^{-\e}$-maximizer of $\g^i$ with respect to $P_k$. By Proposition \ref{approx length prop}, we have $L_{\eta^{-\e}}(\g^i) = \lim_{k \to \infty}L_{\eta^{-\e}}(\l^i_k)$. Fix $k$ large enough so that $L_{\eta^{-\e}}(\g^i) \geq L_{\eta^{-\e}}(\l^i_k) - \e/l$ for each $i$. Using  Lemma \ref{L and L_h lem}, we have


\begin{align*}
L(\g) \,=\, \sum_{i = 1}^l L(\g^i)
\,&\geq\, \sum_{i = 1}^l \bigg(L_{\eta^{-\e}}(\g^i) - CL_h(\g^i)\sqrt{\e} \bigg)
\\
&=\, \sum_{i = 1}^l L_{\eta^{-\e}}(\g^i) \,-\, CL_h(\g)\sqrt{\e}
\\
&\geq\, \sum_{i = 1}^l \bigg( L_{\eta^{-\e}}(\l^i_k) \,-\, \e/l \bigg) \,-\,CL_h(\g) \sqrt{\e}
\\
&=\,\sum_{i = 1}^l L_{\eta^{-\e}}(\l^i_k) \,-\, \bigg(\e + CL_h(\g)\sqrt{\e}\bigg).
\end{align*}
Recall that $\l^i_k$ is the concatenation of the curves $\l^i_{k,j}$ for $j = 1, \dotsc, k$ where $\l^i_{k,j}$ is the straight line in $B^i_\e$ joining $p^i_{k,j}$ to $q^i_{k,j}$ where
\[
p^i_{k,j} \,=\,\g^i\left(t^{i-1} + \frac{j-1}{k}(t^i -t^{i-1})\right) \:\:\:\:\:\: \text{ and } \:\:\:\:\:\: q^i_{k,j} \,=\, \g^i\left(t^{i-1} + \frac{j}{k}(t^i -t^{i-1})\right).
\]
We extend $\l^i_{k,j}$ just slightly to a new curve $\tilde{\l}^i_{k,j}$ such that $\tilde{\l}^i_{k,j}$ is still the straight line between its end points $\tilde{p}^i_{k,j}$ and $\tilde{q}^i_{k,j}$. Moreover, we choose the points $\tilde{p}^i_{k,j}$ and $\tilde{q}^i_{k,j}$ sufficiently close to $p^i_{k,j}$ and $q^i_{k,j}$, respectively, such that
\begin{align*}
L_{\eta^{-\e}}(\l^i_k) \,&=\, \sum_{j = 1}^k L_{\eta^{-\e}}(\l^i_{k,j}) \,\geq \, \sum_{j =1}^k L_{\eta^{-\e}}(\tilde{\l}^i_{k,j}) - \e/l .
\\
 \end{align*}
Therefore
\begin{align*}
L(\g) \,&\geq\, \sum_{i = 1}^l L_{\eta^{-\e}}(\l^i_k) \,-\, \bigg(\e + CL_h(\g)\sqrt{\e}\bigg)
\\
&\geq\, \sum_{i = 1}^l \bigg(\sum_{j =1}^k L_{\eta^{-\e}}(\tilde{\l}^i_{k,j}) - \e/l   \bigg) \,-\, \bigg(\e + CL_h(\g)\sqrt{\e}\bigg)
\\
&=\, \sum_{i = 1}^l \sum_{j = 1}^k L_{\eta^{-\e}}(\tilde{\l}^i_{k,j}) \,-\, \bigg(2\e + CL_h(\g)\sqrt{\e}\bigg).
\end{align*}
Let $D^i_{k,j} \subset V^i_\e$ denote the diamond 
\[
D^i_{k,j} \,=\, I^+_{\eta^{-\e}}(\tilde{p}^i_{k,j}, V^i_\e) \cap I^-_{\eta^{-\e}}(\tilde{q}^i_{k,j}, V^i_\e).
\]
Let $C^i_{k,j} \subset \R$ denote the closed interval 
\[
C^i_{k,j} \,=\, \left[t^{i-1} + \frac{j-1}{k}(t^i -t^{i-1}),\: t^{i-1} + \frac{j}{k}(t^i -t^{i-1})\right].
\]
By uniform convergence of the $\g_n$, there is an integer $N$ such that $n \geq N$ implies $\g_n|C^i_{k,j}$ is contained in $D^i_{k,j}$ for all $i = 1, \dotsc, l$ and all $j = 1, \dotsc, k$. Since $\tilde{\l}^i_{k,j}$ is an $\eta^{-\e}$-maximizer, the proof of (2) from Proposition \ref{eta-maximizer prop} shows that
\[
L_{\eta^{-\e}}(\tilde{\l}^i_{k,j}) \,\geq\, L_{\eta^{-\e}} (\g_n|C^i_{k,j}).
\]

\noindent 
Therefore
\[
L(\g) \,\geq\, \sum_{i = 1}^l \sum_{j=1}^k L_{\eta^{-\e}}(\g_n|C^i_{k,j})  \,-\, \bigg(2\e + CL_h(\g)\sqrt{\e}\bigg).
\]
\noindent Using Lemma \ref{L and L_h lem}, we have
\[
\sum_{i = 1}^l \sum_{j=1}^k L_{\eta^{-\e}}(\g_n|C^i_{k,j}) \,\geq\, L(\g_n) \,-\, CL_h(\g)\sqrt{\e}.
\]
Thus, for these $n \geq N$, we have
\[
L(\g) \,\geq\, L(\g_n) - \bigg(2\e + 2CL_h(\g) \sqrt{\e} \bigg).
\]
Since $\e$ was arbitrary, the result follows.
\qed

\medskip
\medskip

\medskip
\medskip

\noindent\underline{\emph{Proof of Proposition \emph{\ref{length func prop}}}}:
\medskip

 Since $\g\big([a,b]\big)$ is compact, we can cover it by finitely many neighborhoods $U^1, U^2 \dotsc, U^l$ given from Lemma \ref{lem for upper semi prop}. We order these sets in the following way: There exists some set which contains $\g(a)$, say $U^1$. Define $s^1$ via $s^1 = \sup \{ t \mid \g\big([a,t)\big) \subset U^1\}$. Either $s^1 = b$ in which case we take $l = 1$ and stop. Otherwise $\g(s^1) \notin U^1$, and so there exists some set which contains $\g(s^1)$, say $U^2$. Define $s^2$ via $s^2 = \sup \{ t \mid \g\big([s^1,t)\big) \subset U^2\}$. Either $s^2 = b$ in which case we take $l = 2$ and stop. Otherwise $\g(s^2) \notin U^2$, and so there exists some set which contains $\g(s^2)$, say $U^3$. And so on until $s^{l} = b \in U^l$.  
  For $i = 1, \dotsc, l-1$, choose $t^i < s^i$  such that $\g(t^i) \in U^{i} \cap U^{i+1}$. Define $\g^1 = \g|_{[a, t^1]}$ and $\g^2 = \g|_{[t^1, t^2]}$ and so on until $\g^l = \g|_{[t^{l-1}, b]}$. Then $\g^i \subset U^i$ for all $i = 1, \dotsc, l$. For each $n$, define $\g^i_n$ with the same domain restriction as $\g^i$. Then $\g^i_n$ converges uniformly to $\g^i$. 
 
 Fix $\e > 0$. By Lemma \ref{lem for upper semi prop}, for each $i$, there exists an $N_i$ such that $n \geq N_i$ implies $L(\g^i) \geq L(\g^i_n) - \e/l$. Let $N = \max\{N_1, \dotsc, N_l\}$. Then for $n \geq N$, we have 
 \[
 \pushQED{\qed}
 L(\g) \,=\, \sum_{i =1}^l L(\g^i) \,\geq\, \sum_{i =1}^l \bigg(L(\g^i_n) - \e/l \bigg) \,=\, L(\g_n) -\e. \qedhere
\popQED 
 \]

\medskip
\medskip


\section{Bubbling spacetimes}

\subsection{Bubbling sets and causally plain spacetimes}\label{bubbling set section}

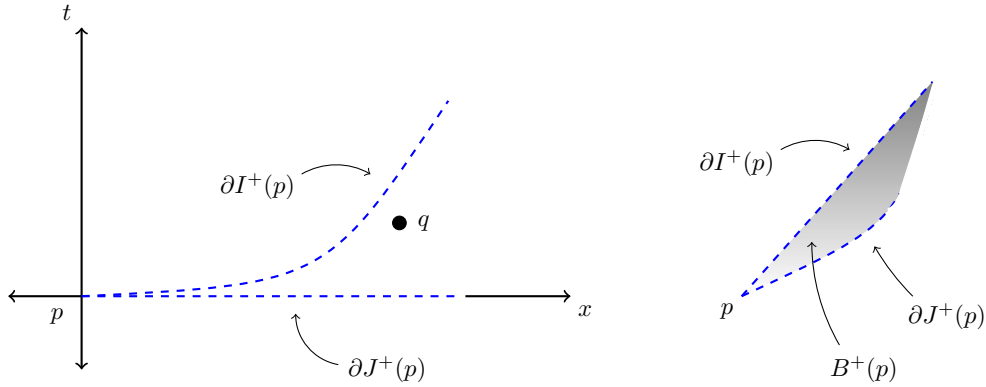
\begin{figure}[h]
\[
\begin{tikzpicture}[scale = 0.65]

\draw [<->,thick] (-5.5,-1.5) -- (-5.5,5.5);
\draw [->,thick] (2.35,0) -- (4.5,0);
\draw[<-, thick] (-7,0) -- (-5.5,0);

\draw (-6,-0.4) node [scale = .85] {$p$};


\draw [dashed, thick, blue] (-5.5,0) .. controls (-0.5,.25) .. (2,4);

\draw [dashed, thick, blue](-5.5,0) -- (2.25,0);

\draw [->] (-1,2.5) arc [start angle=120, end angle=60, radius=40pt];
\draw (-1.9,2.25) node [scale =.85] {$\pd I^+(p)$};

\draw [->] (-0.25,-1.4) arc [start angle=-100, end angle=-180, radius=30pt];
\draw (0.75,-1.5) node [scale =.85] {$\pd J^+(p)$};

\node [scale = .50] [circle, draw, fill = black] at (1, 1.5)  {};
\draw (1.5,1.5) node [scale =.85] {$q$};


\draw (-5.8,5.8) node [scale = .85] {$t$};

\draw (4.8,-0.3) node [scale = .85] {$x$};


\shadedraw [dashed, thick, blue]
(8,0) .. controls (11.25,1.5)  .. (12, 4.5);

\draw [ultra thick, white]
(11, 1.75) .. controls (11.25, 1.90) .. (11.99, 4.5);

\draw [ultra thick, white]
(11, 1.75) .. controls (11.30, 1.85) .. (11.99, 4.5);

\draw [thick, white]
(11.235,2.1) -- (12,4.5);

\draw [dashed, thick, blue]
(8,0) -- (11.9,4.3875);

\draw (7.7,-0.3) node [scale = .85] 
{$p$};

\draw [->] (8.8,3) arc [start angle=120, end angle=60, radius=40pt];
\draw (7.9,2.725) node [scale =.85] {$\pd I^+(p)$};

\draw [->] (11.5,0) arc [start angle=-135, end angle=-150, radius=140pt];
\draw (12.2,-.4) node [scale =.85] {$\pd J^+(p)$};

\draw [->] (10,-1) arc [start angle=-150, end angle=-180,
radius=120pt];
\draw (10.5,-1.5) node [scale =.85] {$B^+(p)$};


\end{tikzpicture}
\]
\captionsetup{format=hang}
\caption{\small{Left: The $C^0$ spacetime from our motivating example. Here $q \in B^+(p)$. Right: A possible bubbling set in a $C^0$ spacetime. Here the shaded region represents the open set $B^+(p)$.}} \label{bubble fig}
\end{figure}

We begin with a motivating example.  Consider the spacetime $(M,g)$ where
\[
M\, = \,(-1,1) \times \R, \:\:\:\:  \:\:\:\:
g \,=\, -dt^2 -a(t)\,dt dx + b(t) \, dx^2
\]
and 
\begin{align*}
a(t) \,&= \,2\big(1 - |t|^{1/2}\big)
\\
b(t) \, &= \, |t|^{1/2}\big(2 - |t|^{1/2}\big).
\end{align*}
Since $\frac{d}{dt} \sqrt{t} = 1/( 2\sqrt{t})$, the metric components $a(t)$ and $b(t)$ are not $C^1$. In fact they are not even Lipschitz. If $\g$ is a curve beginning at the origin, then $\g(x) = \big(t(x),\, x\big)$ will be null when 
\[
t'(x) \, =\, |t|^{1/2}.
\]
Since $|t|^{1/2}$ is continuous, we are guaranteed the existence of solutions. However it is not Lipschitz, so we are not guaranteed uniqueness. Indeed the solutions for the initial condition $t(0) = 0$ are given by the bifurcating family
\[   t_{c}(x) = \left\{
\begin{array}{ll}
      0 & \text{ for } \: 0 \,\leq \, x \,\leq\, c 
      \\ 
      \frac{1}{2}(x - c)^2  & \text{ for } \: c \, \leq \,  x. \\
\end{array} 
\right. \]
If we let $p = (0,0) \in M$ denote the origin, then this example demonstrates the proper inclusion:
\[
I^+(p)  \, \subsetneq \, \text{int}\big[ J^+(p)\big].
\]
See Figure \ref{bubble fig}. 

\medskip
\medskip

\medskip
\medskip

\begin{Def}
\emph{
Let $(M,g)$ be a $C^0$ spacetime. Given a set $S$ within a neighborhood $U$, we define the future \emph{bubbling set} of $S$ within $U$ as the open set
\[
B^+(S, U) = \text{int}\big[ J^+(S,  U)\big] \, \setminus \, \ov{I^+(S, U)}.
\]
}
\end{Def}

\medskip
\medskip

\noindent\emph{Remark.} Past bubbling sets are defined time-dually. When $U = M$, we will simply write $B^+(S)$. When $S = \{p\}$ we will simply write $B^+(p,U)$.

\medskip
\medskip

Since bubbling sets are unfamiliar (and hence undesirable), we set out to establish sufficient conditions which will guarantee they are empty.

\medskip
\medskip

\begin{Def} \emph{Let $(M,g)$ be a $C^0$ spacetime.}
\:
\begin{itemize}
\item[-]\emph{$(M,g)$ is called \emph{causally plain} if $B^+(p) = \emptyset$ for all $p$.
}
\item[-]\emph{$(M,g)$ satisfies the \emph{push-up property} if  $I^+\big(J^+(p)\big) = I^+(p)$ for all $p$. }
\end{itemize}
\end{Def}

\medskip
 
\begin{prop}
If $(M,g)$ satisfies the push-up property, then $(M,g)$ is causally plain.
\end{prop}
 
\proof
Fix $q \in \text{int}\big[J^+(p)\big]$. Then there is a neighborhood $U \subset \text{int}\big[J^+(p)\big]$ about $q$. Therefore there is a causal curve from $p$ to a point $q' \in I^-(q,U)$. Thus  $q \in I^+(p)$ by the push-up property. Hence $B^+(p) = \emptyset$.
\qed 
 
\medskip
\medskip 
 
  In Appendix \ref{C2 appendix} we demonstrate that normal neighborhoods can be used to show that $C^2$ spacetimes satisfy the push-up property. Therefore they are causally plain. The motivating $C^0$ spacetime from the beginning of this section demonstrates that spacetimes with regularity less than $C^1$ are not causally plain. Therefore a natural question to ask is: are $C^1$ spacetimes causally plain? The answer is yes. In fact Lipschitz is sufficient.

\medskip
\medskip

\begin{Def}
\emph{
A \emph{Lipschitz} spacetime $(M,g)$ is one such that the components of the metric $g_{\mu\nu}$ in any coordinate system are locally Lipschitz functions. 
}
\end{Def}

\medskip
\medskip

The following proof is inspired by \cite[Lemma 1.15]{ChrusGrant}.

\medskip
\medskip

\begin{thm}\label{lip implies plain thm}
If $(M,g)$ is Lipschitz, then $(M,g)$ satisfies the push-up property. Hence Lipschitz spacetimes are causally plain.
\end{thm}

\proof
Let $(M,g)$ be a Lipschitz spacetime. Let $\g \colon [0,b] \to M$ be a causal parameterized by $h$-arclength. We will construct a timelike curve $\l \colon [0,b] \to M$ with $\l(0) = \g(0)$ and  such that $\l(b) \in I^+\big(\g(b)\big)$ and $\l(b)$ can be made arbitrarily close to $\g(b)$. Since $I^-$ is open, this will prove the push-up property.
 
Since $(M,g)$ is time-oriented, there is a $C^1$ future-directed timelike vector field $T$ on $M$.  Let $c_p$ denote the integral curve of $T$ through $p = c_p(0)$. Let $f \colon [0,b] \to \R$ be a (soon to be determined) continuous function with $f(0) = 0$ and $f(t) \geq 0$. Let $\e > 0$. For each $t \in [0,b]$, we define $\l(t) = c_{\g(t)}\big(\e f(t)\big)$. Since integral curves are unique, this uniquely defines $\l$.

Since $T \neq 0$, given any point $p$, we can construct a neighborhood $U$ around $p$ with compact closure such that $T = \pd/\pd x^0$ within this neighborhood (see Lemma 1.57 in \cite{ON}). Since $[0,b]$ is compact, we can cover $\g$ be finitely many such neighborhoods $\phi_\a \colon U_\a \to M$ for finite $\a$. By choosing $\e$ small enough, we can ensure $\l \subset \cup_\a U_\a$. In a particular neighborhood $U_\a$, we have by construction $T = T^\mu_\a \pd /\pd x_\a^\mu$ where $T^0_\a = 1$ and $T^i_\a = 0$. Therefore we have the following coordinate expression for $\l(t) \in U_\a$ 
\[
\l^\mu_\a(t) \, = \, \g^\mu_\a(t) \, + \, \e f(t) T^\mu_\a.
\]
This shows that $\l$ is a locally Lipschitz curve. Since $(M,g)$ is Lipschitz and there are only finitely many $\a$, there is a $\Lambda > 0$ such that for any $\a$ and any $x,y \in U_\a$, we have $|g_{\mu\nu}(x) - g_{\mu\nu}(y)| \leq \Lambda |\phi_\a(x) - \phi_\a(y)|$ where $|\cdot|$ denotes the standard Euclidean norm. 

%
%
 
Let $X = \l'$. Omitting the subscript $\a$, we simply write $\l' = X^\mu \pd_\mu$. To utilize the Lipschitz assumption, we separate in any neighborhood $U_\a$
\begin{align*}
g\big(\l'(t), \l'(t)\big) \, &= \big[g_{\mu\nu} \circ \l(t)\big]X^{\mu}(t)X^\nu(t)
\\
&=\, \big[g_{\mu\nu}\circ\l(t) - g_{\mu\nu}\circ \g(t) \big]X^{\mu}(t)X^\nu(t) \,+\, \big[g_{\mu\nu}\circ \g(t)\big] X^{\mu}(t)X^\nu(t)
\end{align*}
Plugging $X^\mu = (\g^\mu)' + \e f'T^\mu$ into the last term above yields
\begin{align*}
g(\l', \l') \,&=\, \big[g_{\mu\nu} \circ \l - g_{\mu\nu}\circ \g\big]X^\mu X^\nu \,+\, g(\g', \g') \,+\, 2\e f' g(\g',T) \,+\, \e^2|f'|^2(g_{00}\circ \g)
\\
\,&\leq\, \big[g_{\mu\nu} \circ \l - g_{\mu\nu}\circ \g\big]X^\mu X^\nu \,+\, 2\e f' g(\g',T).
\end{align*}
The Lipschitz assumption implies that for all $\a$
\[|g_{\mu\nu} \circ \l - g_{\mu\nu} \circ \g| \,\leq\, \Lambda |\phi_\a\circ\l - \phi_\a \circ\g| \,=\, \Lambda \e f.
\]
Therefore in each of the neighborhoods $U_\a$, we have
\[
\big|\big(g_{\mu\nu} \circ \l - g_{\mu\nu}\circ \g\big)X^\mu X^\nu\big| \,\leq\, \e\Lambda f \sum_{\mu, \nu}|X^\mu X^\nu|.
\]
Write $Y = \g'$. We have for each $\mu$
\begin{align*}
|X^\mu|^2 \, \,&= \, |Y^\mu \,+\, \e f'T^\mu|^2 
\\
 &\leq \, |Y^\mu|^2 \,+\, 2\e|Y^\mu||f'| \,+\, \e^2|f'|^2 
 \\
 &\leq\,  C^2 \,+\, 2\e C|f'| \,+\, \e^2|f'|^2.
\end{align*}
The last inequality follows since compactness of  $[0,b]$ along with Proposition \ref{locally lipschitz components prop} implies that there is a constant $C > 0$ (independent of $\a$) such that $|Y^\mu| \leq C$ for all $\mu$.

  Letting $N = n+1$ denote the dimension of the spacetime, we have
\begin{align*}
g(\l',\l') \,&\leq\, \big[g_{\mu\nu} \circ \l - g_{\mu\nu}\circ \g\big]X^\mu X^\nu \,+\, 2\e f' g(\g',T)
\\
\,&\leq\, \e\Lambda fN^2\big(C^2 \,+\, 2\e C|f'| \,+\, \e^2|f'|^2\big) \,+\, 2 \e f' g(\g', T)
\end{align*}
Now we put a bound on the sum of the first and fourth terms. Define the continuous function $r \colon [0,b] \to \R$ via $r(t) = -g\big(\g'(t), T\big)$. Consider the initial value problem $\Lambda C^2N^2 f - 2f'r = -2$ with $f(0) = 0$. A solution is $f(t) = \frac{1}{D}\big[e^{\int_0^t\frac{D}{r(s)}ds} -1\big]$ where $D = \frac{1}{2}\Lambda C^2N^2$.  With this choice of $f$, the first and fourth terms combine to give 
\[
g(\l', \l') \,\leq\, \e\Lambda N^2f\big( 2\e C |f'| + \e^2 |f'|^2\big) -2\e.
\]
Claim: $1/r$ is bounded almost everywhere in $[0,b]$. Assuming this claim for now, we have $f$ is continuous and so it is bounded in $[0,b]$. And since $f' =  (Df+1)/r$, we also have $|f'|$ is bounded almost everywhere in $[0,b]$. Therefore we can choose $\e$ small enough so that 
 \[
g(\l', \l') \,\leq\, -\e
\]
almost everywhere. Thus $\l$ is a timelike curve. Since $r \geq 0$, we have $f \geq 0$. Therefore  $\l(b) \in I^+\big(\g(b)\big)$ and we can make $\l(b)$ arbitrarily close to $\g(b)$ by choosing $\e$ sufficiently small. This proves the theorem once we prove the claim.
 
 Now we prove the claim. Showing $1/r$ is bounded almost everywhere in $[0,b]$ is equivalent to showing that there exists a $c > 0$ such that $-g(\g', T) \geq c$ almost everywhere in $[0,b]$.  Hence it suffices to show $c > 0$ where
 \[
c \,=\, \inf \,\{-g(Z, T) \mid p \in \cup_a U_\a, \:Z \in T_pM, \: g(Z, T) < 0,\:  h(Z,Z) = 1\}.
\]
Since $\cup_\a U_\a$ has compact closure, the same argument used in the proof of Proposition \ref{causal limit curve prop} shows that $c > 0$. This proves the claim.
\qed


\medskip
\medskip


\subsection{Trapped sets in $C^0$ spacetimes}\label{trapped set section}
 
\medskip

Trapped sets play a prominent role in $C^2$ causal theory where they are used to prove the existence of singularities in a spacetime (i.e incomplete geodesics). The most notable example of this is Penrose's original singularity theorem \cite{Penrose, Senovilla}. In this section we offer a definition for trapped sets in $C^0$ spacetimes and prove a $C^0$ version of Penrose's theorem: if $(M,g)$ has a noncompact Cauchy surface, then there are no trapped sets in $M$.

\medskip
\medskip

\begin{Def} \emph{Let $(M,g)$ be a $C^0$ spacetime.}
\begin{itemize}
\item[-] \emph{$F$ is a \emph{future set} if $I^+(F) \subset F$.}

\item[-] \emph{$\S$ is \emph{future trapped} if there is a nonempty future set $F \subset J^+(\S)$ such that $\pd F$ is compact. }
\end{itemize}
\end{Def}

\medskip
\medskip

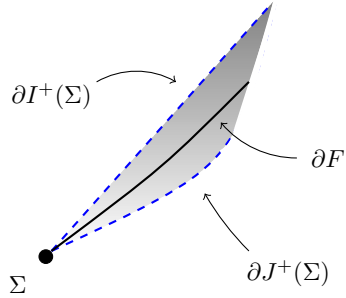
\begin{figure}[h]
\[
\begin{tikzpicture}[scale = 0.75]


\shadedraw [dashed, thick, blue]
(8,0) .. controls (11.25,1.5)  .. (12, 4.5);

\draw [ultra thick, white]
(11, 1.75) .. controls (11.25, 1.90) .. (11.99, 4.5);

\draw [ultra thick, white]
(11, 1.75) .. controls (11.30, 1.85) .. (11.99, 4.5);

\draw [thick, white]
(11.235,2.1) -- (12,4.5);

\draw [dashed, thick, blue]
(8,0) -- (11.9,4.3875);

\draw [thick, black]
(8,0) .. controls (10,1.5)  .. (11.5, 3);

\draw [->] (12.35,1.65) arc [start angle=-90, end angle=-140,
radius=50pt];
\draw (12.9,1.65) node [scale =.85] {$\pd F$};

\draw (7.4,-0.6) node [scale = .85] 
{$\S$};

\draw [->] (8.8,3) arc [start angle=120, end angle=60, radius=40pt];
\draw (8,2.75) node [scale =.85] {$\pd I^+(\S)$};

\draw [->] (11.5,0) arc [start angle=-135, end angle=-150, radius=140pt];
\draw (12.2,-.4) node [scale =.85] {$\pd J^+(\S)$};

\node [scale = .5] [circle, draw, fill = black] at (7.9,-.1)  {};


\end{tikzpicture}
\]
\captionsetup{format=hang}
\caption{\small{A future set $F \subset  J^+(\S)$. If $\pd F$ is compact, then $\S$ is future trapped.}} \label{trapped set fig}
\end{figure}

Examples of future sets are $I^+(\S)$ and $J^+(\S)$. In bubbling spacetimes, the boundaries of these future sets may not be equal: $\pd I^+(\S) \neq \pd J^+(\S)$, and so there can be a future set with boundary $\pd F$ that lies between them. See Figure \ref{trapped set fig}.

\medskip
\medskip

\medskip
\medskip

\begin{prop}
If $F$ is a future set, then $\pd F$ is achronal and has empty edge.
\end{prop}

\proof
We first show $\pd F$ is achronal. Fix $p, q \in \pd F$ and suppose there is a timelike curve from $p$ to $q$. Then  $I^-(q)$ is an open set containing  $p \in \pd F$. Hence it also contains a point $r \in F$. Therefore $q \in I^+(r)$. But the definition of a future set implies $I^+(r) \subset F$. Hence $q \in \text{int}[F]$ which contradicts the assumption $q \in \pd F$. 

Now we show $\pd F$ has no edge points. Fix $p_0 \in \pd F$. Let $p \in I^-(p_0)$ and $q \in I^+(p_0)$ and let $\g \colon [a,b] \to M$ be a timelike curve from $p$ to $q$. Since $q \in I^+(p_0)$, we have $q \in \text{int}[F]$ by definition of a future set. Since $\pd F$ is achronal, we have $I^-(p_0) \subset M \setminus F$. Hence $p \in \text{int}[M\setminus F]$. Define
\[
t_* \,=\, \inf \big\{t \in [a,b] \mid \g\big((t,b]\big) \subset \text{int}[F]\big\}.
\]
Since $\text{int}[F]$ is open, we have $t_* < b$. Likewise $p \in \text{int}[M \setminus F]$ implies $t_* > a$. Since $\g(t_*)$ is an accumulation point of $\text{int}[F]$, we have $\g$ intersects $\pd F$: 
\[
\pushQED{\qed}
\g(t_*) \,\in\, \pd \,\text{int}[F] \,=\, \ov{\text{int}[F]} \setminus \text{int}[F] \,\subset\, \ov{F} \setminus \text{int}[F] \,=\, \pd F. \qedhere
\popQED
\]

\medskip
\medskip

\begin{cor}\label{future set cor}
If $F$ is a nonempty future set, then $\pd F$ is a $C^0$ hypersurface.
\end{cor}

\proof
This follows from Theorem \ref{achronal edge thm}.
\qed

\medskip
\medskip

The following proof is a direct analogue of Penrose's original proof \cite{ON}. We include it for the sake of (in)completeness. 

\medskip
\medskip

\begin{thm}[Penrose]\label{Penrose thm}
Let $(M,g)$ be a $C^0$ spacetime with a noncompact Cauchy surface. Then there are no future trapped sets in $M$.
\end{thm}

\proof
Let $S$ be the Cauchy surface. Claim: $S$ is connected. Since $M$ is time-oriented there is a $C^1$ timelike vector field $\tilde{X}$ on $M$. Let $X = \tilde{X}/h(\tilde{X},\tilde{X})^{1/2}$ so that $h(X,X) = 1$. Since maximal integral curves are inextendible as continuous curves, the integral curves of $X$ are inextendible causal curves and are parameterized by $h$-arclength by construction. Let $\g_p \colon \R \to M$ denote the maximal integral curve of $X$ through $p$. Let $\phi \colon M \times \R \to M$ denote the flow of $X$ given by $\phi(p,t) =\g_p(t)$. Let $\phi_S\colon S \times \R \to M$ denote the restriction of $\phi$ to $S \times \R$. Then $\phi_S$ is one-to-one since integral curves don't intersect, and $\phi_S$ is onto since $S$ is a Cauchy surface. Since $S$ is a $C^0$ hypersurface by 
Corollary \ref{Cauchy surface is top sur}, Brouwer's invariance of domain theorem implies $\phi_S$ is a homeomorphism. Let $\pi \colon S \times \R \to S$ denote the natural projection. Put $r = \pi \circ \phi_S^{-1}$. Then $r \colon M \to S$ is a retraction of $M$ onto $S$. Since $M$ is connected, $S = r(M)$ is connected. This proves the claim.

 Seeking a contradiction, suppose $\S$ is future trapped with future set $F$. Let $r_{\pd F} \colon \pd F \to S$ denote the restriction of $r$ to $\pd F$. Since $r_{\pd F}$ is one-to-one and $\pd F$ is a $C^0$ hypersurface, Brouwer's invariance of domain theorem implies $r_{\pd F}$ is a homeomorphism of $\pd F$ onto an open subset of $S$. Since $\pd F$ is compact, $r_{\pd F}(\pd F)$ is closed in $S$. Therefore $r_{\pd F}(\pd F) = S$ since $S$ is connected. But this contradicts $S$ being noncompact.
 \qed

\medskip
\medskip

\noindent\emph{Remark.} It would be interesting to see what conditions on a $C^0$ spacetime would force a future trapped set. For instance in a $C^2$ spacetime we have \cite[Proposition 14.60]{ON}:
\[
\text{Trapped surface + null energy condition +  null completeness \: $\Longrightarrow$ \: future trapped set}
\] 

\medskip
\medskip

\subsection*{Acknowledgments}

The author thanks Greg Galloway, Piotr Chru{\'s}ciel, and Annegret Burtscher for helpful comments and discussions. This material is based upon work supported by the Swedish Research Council under grant no. 2016-06596 while the author was a participant at Institut Mittag-Leffler in Djursholm, Sweden during the Fall semester of 2019.

\newpage

\appendix

\section{Appendices}

\subsection{Differences between $C^0$ and smooth (at least $C^2$) causal theory}\label{C2 appendix}

In this appendix we highlight the main difference between causal theory in smooth (at least $C^2$) spacetimes and causal theory in $C^0$ spacetimes. The goal is to see how the twice-differentiability of the metric is used in $C^2$ causal theory and the difference that arises with $C^0$ metrics. For references on $C^2$ causal theory  one can look at classical sources such as \cite{Wald, ON} or more recent sources such as \cite{Chrus, MinguzziLivRev}.

Let $(M,g)$ be a $C^2$ spacetime. Then there is a unique affine connection $\nabla$ such that $\nabla g = 0$. A curve $\g$ is a \emph{geodesic} if $\nabla_{\g'} \g' = 0$. A consequence of $\nabla g = 0$ is that a geodesic must be either timelike, null, or spacelike.  The equation $\nabla_{\g'}\g' = 0$ is a second order differential equation. Introducing a coordinate system $x^\mu$ and putting $\g^\mu = x^\mu \circ \g$, this differential equation is $\frac{d^2 \g^\mu}{dt^2} + \Gamma^\mu_{\a\b} \frac{d\g^\a}{dt}\frac{d\g^\b}{dt} \,=\, 0.$

Since the metric is $C^2$, the Christoffel symbols are $C^1$. Thus the fundamental existence and uniqueness theorem for differential equations implies a map
$
\exp_p\colon D\subset T_pM \,\to\, M
$
 called the \emph{exponential map} given by $\exp_p(X) = \g(1)$ where $\g \colon [0,1] \to M$ is the unique geodesic satisfying $\g(0) = p$ and $\g'(0) = X$. The set $D \subset T_pM$ is defined by requiring $X \in D$ implies $\g(1)$ is defined. The derivative of $\exp_p$ at the origin is just the identity, so by the inverse function theorem, for any point $p \in M$ there is an open set $D \subset T_pM$ such that $\exp_p\colon D \to M$ is a diffeomorphism onto its image $U = \exp_p(D)$. In this case $U$ is called a \emph{normal neighborhood}.   For these open sets, we have the following characterizations \cite{Chrus}:
\begin{enumerate}

\item[-] points in $I^+(p,U)$ correspond to future-directed timelike vectors in $D \subset T_pM$

\item[-] points in $\pd I^+(p,U)$ correspond to future-directed null vectors in $D \subset T_pM$ 

\end{enumerate}

\begin{figure}[h]
\[
\begin{tikzpicture}[scale = 0.65]

\draw [<->,thick] (-7,-3.5) -- (-7,3.5);
\draw [<->,thick] (-11,0) -- (-3,0);

\draw [loosely dashed, thick] (-7,0) circle (75pt);

\draw [dashed, thick, blue](-8.75,1.75) -- (-7,0) -- (-5.25,1.75);
\draw [dashed, thick, blue]
(-8.75,-1.75) -- (-7,0) -- (-5.25,-1.75);

\draw [->, ultra thick] (-7,0) -- (-7.75, 1.75);
\draw (-7.5,2.1) node [scale = .85] {$X$};

\draw [->, ultra thick] (-7,0) -- (-5.80,1.20);
\draw (-5.5,0.5) node [scale = .85] {$Y$};

\draw (-7,-4) node [scale = .85] {$D \, \subset \, T_pM$};


\draw [->, thick] (-1.8,1) -- (.8,1);
\draw (-.5,1.5) node [scale = .85] {$\exp_p$};


\draw [loosely dashed, thick] (3,0) .. controls (4,6) and (9,2) .. (9,-1)
.. controls (6,-4) and (4,-1) .. (3,0);

\draw [dashed, thick, blue]
(6,0) .. controls (5,.5) and (5, 2) .. (4, 2.5);

\draw [dashed, thick, blue] (7.25, 1.25) -- (7.5, 2);

\draw [dashed, thick, blue]
(6,0) .. controls (5, -1) .. (4.75, -1.75); 

\draw [dashed, thick, blue]
(6,0) .. controls (7, -1) .. (8, -1.75);

\draw (6, -0.5) node [scale = .85] {$p$};

\draw [ultra thick] (6,0) .. controls (5.25, 1.25) .. (5, 2.25);

\draw [ultra thick] (6,0) .. controls (7.0, 0.6) .. (7.20, 1.15);

\node [scale = .5] [circle, draw, fill = black] at (5,2.25)  {};
\node [scale = .5] [circle, draw, fill = black] at (7.20,1.1)  {};

\draw (5.5, 2.5) node [scale = .85] {$x$};
\draw (7.75, 1.0) node [scale = .85] {$y$};

\draw (6,-4) node [scale = .85] {$U \, \subset \, M$};

\end{tikzpicture}
\]
\captionsetup{format=hang}
\caption{\small{The point $x = \exp_p(X)$ lies on the timelike geodesic generated by  the timelike vector $X$. The point $y = \exp_p(Y)$ lies on the null geodesic generated by the null vector $Y$.}}
\label{convex fig}
\end{figure}
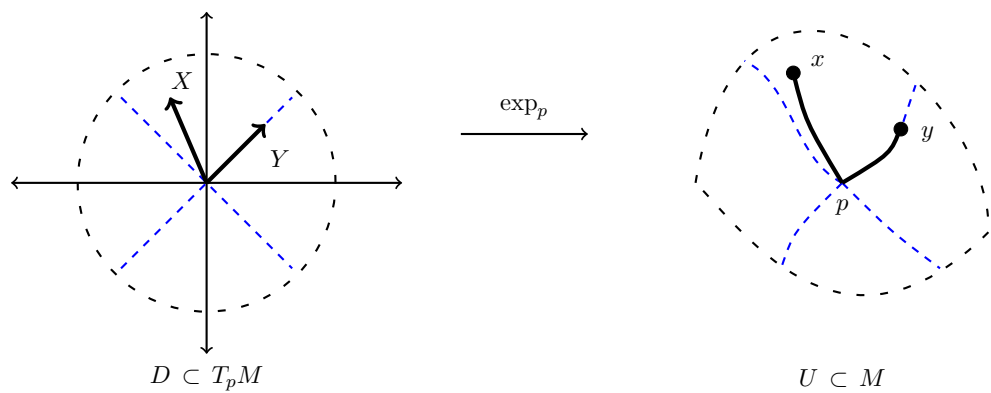

 If $\g$ is a causal curve from $p$ to $q$ and $\l$ is a timelike curve from $q$ to $r$, then using a finite number of normal neighborhoods and the properties above,  we can deform the concatenation of $\g$ and $\l$ into a timelike curve from $p$ to $r$ \cite{Chrus, ON}. This proves the push-up property.

\medskip

\begin{prop}[Push-up property]
Let $(M,g)$ be a $C^2$ spacetime. Then
\[I^+\big(J^+(p)\big) \, = \, I^+(p).
\]
\end{prop}

The push-up property implies:

\medskip
\medskip

\begin{prop}\label{int = I appendix prop} Let $(M,g)$ be a $C^2$ spacetime. Then
\[\emph{\text{int}}\big[J^+(p)\big] \, = \, I^+(p).
\]
\end{prop}

\proof
Fix $q \in \text{int}\big[J^+(p)\big]$. Then there is a normal neighborhood $U \subset \text{int}\big[J^+(p)\big]$ about $q$. Therefore there is a causal curve from $p$ to a point $q' \in I^-(q,U)$. Thus  $q \in I^+(p)$ by the push-up property.
\qed

\medskip
\medskip

It was shown in \cite{ChrusGrant} that Proposition \ref{int = I appendix prop} need not hold for $C^0$ spacetimes. See the \linebreak example in the beginning of section \ref{bubbling set section}. There can be nonempty \emph{bubbling sets} in $C^0$ \linebreak spacetimes. These are the open sets 
\[
B^+(p) \,=\, \text{int}\big[J^+(p)\big] \,\setminus\, \ov{I^+(p)}.
\]
For $C^2$ spacetimes $B^+(p) = \emptyset$ for all $p$ by Proposition \ref{int = I appendix prop}. This highlights the main difference between $C^2$ and $C^0$ causal theory.

\medskip
\medskip

\begin{figure}[h]
\[
\begin{tikzpicture}[scale = 0.75]

\shadedraw [dashed, thick, blue]
(8,0) .. controls (11.25,1.5)  .. (12, 4.5);

\draw [ultra thick, white]
(11, 1.75) .. controls (11.25, 1.90) .. (11.99, 4.5);

\draw [ultra thick, white]
(11, 1.75) .. controls (11.30, 1.85) .. (11.99, 4.5);

\draw [thick, white]
(11.235,2.1) -- (12,4.5);

\draw [dashed, thick, blue]
(8,0) -- (11.9,4.3875);

\draw (7.7,-0.3) node [scale = .85] 
{$p$};

\draw [->] (8.8,3) arc [start angle=120, end angle=60, radius=40pt];
\draw (8,2.75) node [scale =.85] {$\pd I^+(p)$};

\draw [->] (11.5,0) arc [start angle=-135, end angle=-150, radius=140pt];
\draw (12.2,-.4) node [scale =.85] {$\pd J^+(p)$};

\draw [->] (10,-1) arc [start angle=-150, end angle=-180,
radius=120pt];
\draw (10.5,-1.5) node [scale =.85] {$B^+(p)$};

\end{tikzpicture}
\]
\captionsetup{format=hang}
\caption{\small{A nonempty bubbling set $B^+(p)$ in a $C^0$ spacetime. For $C^2$ spacetimes $B^+(p) = \emptyset$ for all points. For $C^0$ spacetimes this begs the question: what should one take as the lightcone? }}\label{bubbling fig in appendix}
\end{figure}
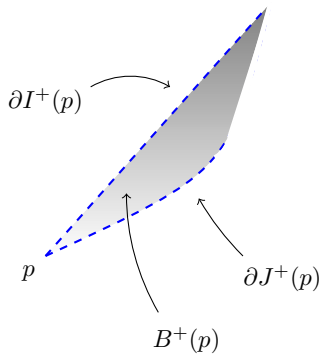

\newpage

\medskip

\subsection{Properties of locally Lipschitz curves}\label{locally Lipschitz appendix}

In Definition \ref{causal curve def} we defined causal and timelike curves via locally Lipschitz curves. In this section we establish the properties of locally Lipschitz curves. These curves are defined via a complete Riemannian metric $h$. Therefore we first show that if $(M,g)$ is a $C^0$ spacetime, then there is a complete Riemannian metric $h$ on $M$.

\medskip

\begin{prop}\label{h existence prop}
Let $M$ be a smooth manifold which is connected, Hausdorff, and second-countable. Then there is a smooth complete Riemannian metric $h$ on $M$.
\end{prop}

\proof
We could construct $h$ directly via a partition of unity as in \cite[Lemma 11.1]{Hans}, but also pointed out in \cite{Hans} is another argument using the Hopf-Rinow and the Whitney embedding theorems (the latter of course still requires a partition of unity argument).

Since $M$ is smooth, Hausdorff, and second-countable, we can apply the Whitney embedding theorem \cite{Lee} to obtain a smooth proper embedding $f \colon M \to \R^N$.  By pulling back the Euclidean metric onto $M$, we have a smooth Riemannian manifold $(M,h)$. Let $d_h$ be the distance function on $M$ induced by $h$. Since $f$ is proper, any closed set in $M$ maps to a closed subset of $\R^N$. Therefore any closed and bounded subset of $(M,d_h)$ will be a closed and bounded subset within $f(M) \subset \R^N$ which is compact by the Heine-Borel theorem. Since $M$ is connected, $(M,h)$ is complete by the Hopf-Rinow theorem.
\qed

\medskip
\medskip

Fix a $C^0$ spacetime $(M,g)$ and a complete Riemannian metric $h$ on $M$.
Let $I \subset \R$ be an interval (i.e. any connected subset of $\R$ with nonempty interior). A \emph{locally Lipschitz} curve $\g \colon I \to M$ is a continuous function such that for any compact $K \subset I$, there is a constant $C$ such that for any $a,b \in K$, we have
\[
d_h\big(\g(a), \g(b)\big) \,\leq\, C|b - a|
\] 
where $d_h$ is the Riemannian distance function associated with $h$.

Of course one normally works with smooth or piecewise smooth curves in a spacetime. That these curves are locally Lipschitz follows from the next proposition.

\medskip
\medskip

\begin{prop}\label{C1 is loc Lip prop}
If $\g \colon I \to M$ is $C^1$, then $\g$ is locally Lipschitz.
\end{prop}

\proof
Let $K \subset I$ be compact. First suppose $\g(K) \subset B$ where $\phi \colon B \to \R^{n+1}$ is a coordinate chart and $\phi(B)$ is an open Euclidean ball in $\R^{n+1}$ with finite radius. Hence $B$ has compact closure. Let $\tilde{\g} = \phi \circ \g$. We first show there is a constant $C > 0$ such that 
\[
d_{h}\big(\g(a), \g(b)\big) \,\leq\, C|\tilde{\g}(a) - \tilde{\g}(b)| \:\:\:\:\:\:\:\: \text{for all } a,b \in K.
 \]
  Define $C$ by
 \[
1/C^2 \,=\, \inf \,\{\delta_{\mu\nu}X^\mu X^\nu \mid p \in B,\: X \in T_pM,\: h(X,X) = 1\}.
\]
Since $B$ has compact closure, we have $1/C^2 > 0$. Then for all $p \in B$ and $X \in T_pM$, we have
\[
h(X,X) \,\leq\, C^2 \delta_{\mu\nu}X^\mu X^\nu.
\]
Fix $a,b \in K$. Let $\s \colon [0,1] \to B$ denote the straight line joining $\g(a)$ to $\g(b)$. That is $\tilde{\s}(t) = \phi \circ \s(t) = t\tilde{\g}(b) + (1-t)	\tilde{\g}(a)$. Let $X = \s'$. Hence $X^\mu = \g^\mu(b) - \g^\mu(a)$. Then
\[
|\tilde{\g}(b) - \tilde{\g}(a)| \, = \, \sqrt{\delta_{\mu\nu}X^\mu X^\nu} \,=\, \int_0^1 \sqrt{\delta_{\mu\nu}X^\mu X^\nu} \,\geq \, \frac{1}{C}\int_0^1 \sqrt{h(\s', \s')} \,\geq\, \frac{1}{C}d_{h}\big(\g(a), \g(b)\big).
\]
Therefore $d_{h}\big(\g(a), \g(b)\big) \leq C|\tilde{\g}(a) - \tilde{\g}(b)|$.

Since $\g$ is $C^1$, there is a constant $c^\mu > 0$ such that $\big|(\g^\mu)'(t)\big| \leq c^\mu$ for all $t \in K$. Let $c = \max_\mu \{c^\mu\}$. Then 
\[
|X^\mu| \,=\, \big|\g^\mu(b) - \g^\mu(a)\big| \,=\, \left|\int_0^1 (\g^\mu)'\big(a + t(b-a)\big)(b-a)dt \right| \,\leq\, c|b-a| .
\]
Therefore
\[
d_h\big(\g(a),\g(b)\big) \,\leq\, C|\tilde{\g}(a) - \tilde{\g}(b)| \,=\, C \sqrt{\delta_{\mu\nu}X^\mu X^\nu} \,\leq\, C\delta_{\mu\nu}\sqrt{X^\mu X^\nu} \,\leq\, C c(n+1) |b-a|.
\]
This proves the proposition when $\g(K) \subset B$. In the general case, we can cover $\g(K)$ by finitely many such balls and then apply the triangle inequality to obtain the result. 
\qed

\medskip
\medskip

Proposition \ref{locally lipschitz components prop} is a partial converse to the previous proposition.

\medskip
\medskip

\noindent\underline{\emph{Proof of Proposition \emph{\ref{locally lipschitz components prop}}}}:
\medskip

Fix $t_0 \in I$. Let $\phi \colon U \to \R^{n+1}$ be a coordinate system around $\g(t_0)$ such that $U$ is a convex open neighborhood (with respect to the Riemannian metric $h$) with compact closure. Define $c$ via
 \[
1/c^2 \,=\, \inf \,\{h(X,X) \mid p \in U,\: X \in T_pM,\: \delta_{\mu\nu}X^\mu X^\nu = 1\}.
\]
Since $U$ has compact closure, we have $1/c^2 > 0$. Then for all $p \in U$ and $X \in T_pM$, we have
\[
\delta_{\mu\nu}X^\mu X^\nu \,\leq\, c^2h(X,X).
\]
Let $K \subset I$ be compact with $\g(K) \subset U$. Fix $a,b \in K$. Since $\g$ is locally Lipschitz, there is a constant $C > 0$ (independent of $a$ and $b$) such that $d_h\big(\g(a), \g(b)\big) \leq C|b-a|$.  Since $U$ is convex, there is a minimizing $h$-geodesic $\s \subset U$ joining $\g(a)$ to $\g(b)$. Let $X = \s'$. Write $\tilde{\g} = \phi \circ \g$. Then
\[
d_h\big(\g(a), \g(b)\big) \,=\, \int \sqrt{h(\s', \s')} \,\geq\, \frac{1}{c}\int \sqrt{\delta_{\mu\nu}X^\mu X^\nu} \,\geq\, \frac{1}{c}\big|\tilde{\g}(a) - \tilde{\g}(b)\big|
\]
where $|\cdot |$ denotes the Euclidean distance in $\R^{n+1}$. The last inequality follows since the shortest distance between $\tilde{\g}(a)$ and $\tilde{\g}(b)$ with respect to the Euclidean metric $\delta_{\mu\nu}$ is just the straight line. 

For any $\mu$, we trivially have $\big|\g^\mu(b) - \g^\mu(a)\big| \leq \big|\tilde{\g}(a) - \tilde{\g}(b)\big|$. Thus 
\[
\big|\g^\mu(b) - \g^\mu(a)\big| \,\leq\, Cc|b-a|.
\]
Hence the components $\g^\mu$ are Lipschitz functions on $K$. Therefore they are absolutely continuous and hence differentiable almost everywhere with derivative bounded by $Cc$ almost everywhere on $K$. \qed

\medskip
\medskip

Lastly, we show that the definition of locally Lipschitz does not depend on the choice of complete Riemannian metric $h$.  See also \cite{Chrus}.

\medskip
\medskip

\begin{prop}\label{indep of h prop}
Let $h_1$ and $h_2$ be complete Riemannian metrics on $M$. Then $\g \colon I \to M$ is locally Lipschitz with respect to $h_1$ if and only if it is locally Lipschitz with respect to $h_2$.
\end{prop}

\proof
Fix a compact set $K \subset I$. Let $L = \int_K \sqrt{h(\g', \g')}$ denote the $h_1$-arclength of $\g|_K$. Note that $L < \infty$ by Proposition \ref{locally lipschitz components prop}. Set $\mathcal{D} = \bigcup_{t \in K}\ov{B
}_{h_1}\big(\g(t), L\big)$. Here $\ov{B}_{h_1}$ denotes the closed geodesic ball with respect to $h_1$. $\mathcal{D}$ is closed since the compactness of $K$ implies its complement is open, and $\mathcal{D}$ is bounded by $3L$. Therefore $\mathcal{D}$ is compact by the Hopf-Rinow theorem. Define $C$ by
\[
1/C^2 \,=\, \inf \,\{h_1(X,X) \mid p \in \mathcal{D}, \: X \in T_pM,\: h_2(X,X) = 1\}.
\]
Compactness of $\mathcal{D}$ implies $1/C^2 > 0$. Then for all $p \in \mathcal{D}$ and $X \in T_pM$, we have
\[
 h_2(X,X) \,\leq \, C^2 h_1(X,X).
\]
Fix $a, b \in K$. Let $\s$ denote a minimizing $h_1$-geodesic between $\g(a)$ and $\g(b)$. Note that the definition of $\mathcal{D}$ implies $\s \subset \mathcal{D}$. Therefore 
\begin{align*}
d_{h_1}\big(\g(a), \g(b)\big) \, &= \, \int \sqrt{h_1(\s', \s')} \,\geq \, C^{-1}\int \sqrt{h_2(\s', \s')} \,\geq\, C^{-1}d_{h_2}\big(\g(a), \g(b)\big).
\end{align*}

Thus, if $\g$ is locally Lipschitz with respect to $h_1$, then it is locally Lipschitz with respect to $h_2$. Reversing the roles of $h_1$ and $h_2$ gives the reverse implication.
\qed

\subsection{Achronal and edgeless subsets in a spacetime}

Fix a $C^0$ spacetime $(M,g)$. A subset $S \subset M$ is \emph{achronal} if $I^+(S) \cap S = \emptyset$. We say $S$ is \emph{achronal in $U$} if $I^+(S, U) \cap S = \emptyset$. We say $S$ is \emph{locally achronal} if for every $p \in S$, there is an open set $U$ around $p$ such that $S$ is achronal in $U$. The \emph{edge} of an achronal set $S$ is the set of points $p \in \ov{S}$ such that for every neighborhood $U$ of $p$, there is a timelike curve $\g \colon [a,b] \to U$ such that $\g(a) \in I^-(p,U)$, $\g(b) \in I^+(p,U)$, and $\g \cap S = \emptyset$. We say $S$ is \emph{edgeless} if $S$ is disjoint from its edge.

 A subset $S \subset M$ is a \emph{$C^0$ hypersurface} provided for each $p \in S$ there is a neighborhood $U \subset M$ and a homeomorphism $\phi \colon U \to \phi(U) \subset \R^{n+1}$ such that $\phi(U \cap S) = \phi(U) \cap P$ where $P$ is a hyperplane in $\R^{n+1}$.

\medskip
\medskip

\noindent\emph{Remark.} The following theorem shows that a locally achronal and edgeless set is a $C^0$ \linebreak hypersurface, but the proof shows that the conclusion can be strengthened to a locally  Lipschitz hypersurface.
 
 \medskip
 \medskip
 
 \newpage

\begin{thm}\label{achronal edge thm}
Let $S \subset M$ be nonempty. If $S$ is locally achronal and edgeless, then $S$ is a $C^0$ hypersurface.  
\end{thm}

\proof 

 Fix $p \in S$.  Let $\phi \colon U_{3/5} \to \R^{n+1}$ be a  coordinate system around $p$ from Lemma \ref{coord lem}. Recall that $\eta^{3/5}$ has lightcones with `slope' 2. Since $S$ is locally achronal, we can shrink our neighborhood $U_{3/5}$ to a new neighborhood $U$ so that $S$ is achronal in $U$. Moreover, since $p$ is not an edge point of $S$, we can further shrink our neighborhood so that every timelike curve beginning in $I^-(p,U)$ and ending in $I^+(p,U)$ must intersect $S$. 
Choose $\a > 0$ small enough so that the hyperplanes $x^0 = \pm \a$ intersect $\phi(U)$. Let $B = \{(b^1, \dotsc, b^n) \mid \sqrt{\delta_{ij}b^ib^j} < \a/2\}$. By shrinking $U$, we can assume $\phi(U) = (-\a, \a) \times B$. For each $b \in B$, there is a vertical line $\g_b\colon (-\a, \a) \to U$ given by $\g_b(t) = \phi^{-1} \circ (t, b)$. Each $\g_b$ intersects $S$ at exactly one point $\xi(b) \in S$.  This defines an injective map $\xi \colon B \to S$. Consider the composition $\phi \circ \xi \colon B \to \R^{n+1}$. Then $\phi \circ \xi(b) = \big(f(b), b\big)$ for some function $f \colon B \to \R$. 

We will show $f$ is Lipschitz (hence continuous) with a  Lipschitz constant $C = 3$. Seeking a contradiction, suppose there exist points $b,b' \in B$ such that $|f(b) - f(b')| \geq 3 |b-b'|$ where $|\cdot|$ denotes the standard Euclidean norm. Let $\l$ be the straight line joining $\big(f(b), b\big)$ and $\big(f(b'), b'\big)$. Then $\l$ is $\eta^{3/5}$-timelike. Hence it is $g$-timelike. This implies $S$ is not achronal in $U$ which is a contradiction. Thus $f$ is Lipschitz. Then 
\[\psi \,=\, (y^0, y^1, \dotsc, y^n) \,=\, \big(x^0 - f\circ(x^1,\dotsc, x^n), x^1, \dotsc, x^n\big)
\] is a homeomorphism from $U$ onto $\psi(U) \subset \R^{n+1}$ such that $\psi(U \cap S) = \psi(U) \cap P$ where $P$ is the hyperplane $y^0 = 0$. 
\qed

\bibliographystyle{amsplain}

\end{document}